\documentclass[10pt,journal,comsoc]{IEEEtran}
\ifCLASSOPTIONcompsoc
  \usepackage[nocompress]{cite}
\else
  \usepackage{cite}
\fi
\ifCLASSINFOpdf
  \usepackage[pdftex]{graphicx}
  \graphicspath{{../plots/}{../figures/}}
\else
\fi
\usepackage{array}

\usepackage{stfloats}
\usepackage[T1]{fontenc}
\usepackage[noend]{algpseudocode}
\usepackage{subcaption}
\usepackage{algorithm}
\usepackage{booktabs}
\usepackage{comment}
\usepackage[draft]{hyperref}
\usepackage{amsmath,amssymb,amsfonts}
\usepackage{graphicx}
\usepackage{mathtools}
\usepackage{textcomp}
\usepackage{xcolor}
\usepackage{float}
\usepackage{bm}
\usepackage{bbold}
\usepackage[normalem]{ulem}
\usepackage{eso-pic}
\usepackage{soul}
\usepackage{multirow}
\usepackage{enumerate}

\begin{document}
%
\title{RouteNet-Fermi: Network Modeling with \\ Graph Neural Networks}
%
%
%
%

\author{Miquel~Ferriol-Galmés,
        Jordi~Paillisse,
        José Suárez-Varela,
        Krzysztof Rusek, 
        Shihan Xiao, 
        Xiang Shi,
        Xiangle~Cheng,
        Pere~Barlet-Ros,
        and~Albert~Cabellos-Aparicio
\IEEEcompsocitemizethanks{\IEEEcompsocthanksitem M. Ferriol, J. Paillisse, J. Su\'arez-Varela, P. Barlet-Ros and A. Cabellos-Aparicio are with the Barcelona Neural Networking Center, Universitat Politècnica de Catalunya, Spain.\protect\\
K. Rusek is with the AGH University of Science and Technology, Poland.\protect\\
S. Xiao, X. Shi, and X. Cheng are with Huawei Technologies Co., Ltd., China.\protect\\
E-mail: \{miquel.ferriol, jordi.paillisse, jose.suarez-varela, pere.barlet, alberto.cabellos\}@upc.edu, krzysztof.rusek@agh.edu.pl, \{xiaoshihan, shixiang16, chengxiangle1\}@huawei.com}
\thanks{This publication is part of the Spanish I+D+i project TRAINER-A (ref.PID2020-118011GB-C21), funded by MCIN/ AEI/10.13039/501100011033. This work is also partially funded by the Catalan Institution for Research and Advanced Studies (ICREA) and the Secretariat for Universities and Research of the Ministry of Business and Knowledge of the Government of Catalonia and the European Social Fund.}

}

\newcommand{\mpp}[1]{\textcolor{blue}{\textbf{M++: } #1}}
\newcommand{\mmm}[1]{\textcolor{red}{\textbf{M--: } #1}}
\newcommand{\com}[2]{\textcolor{orange}{\textbf{Comment: } [#1] #2}}


\IEEEtitleabstractindextext{%
\begin{abstract}
Network models are an essential block of modern networks. For example, they are widely used in network planning and optimization. However, as networks increase in scale and complexity, some models present limitations, such as the assumption of Markovian traffic in queuing theory models, or the high computational cost of network simulators. Recent advances in machine learning, such as Graph Neural Networks (GNN), are enabling a new generation of network models that are data-driven and can learn complex non-linear behaviors. In this paper, we present RouteNet-Fermi, a custom GNN model that shares the same goals as Queuing Theory, while being considerably more accurate in the presence of realistic traffic models. The proposed model predicts accurately the delay, jitter, and packet loss of a network. We have tested RouteNet-Fermi in networks of increasing size (up to 300 nodes), including samples with mixed traffic profiles ---~e.g., with complex non-Markovian models~--- and arbitrary routing and queue scheduling configurations. Our experimental results show that RouteNet-Fermi achieves similar accuracy as computationally-expensive packet-level simulators and scales accurately to larger networks. Our model produces delay estimates with a mean relative error of 6.24\% when applied to a test dataset of 1,000 samples, including network topologies one order of magnitude larger than those seen during training. Finally, we have also evaluated RouteNet-Fermi with measurements from a physical testbed and packet traces from a real-life network.
\end{abstract}

}

\AddToShipoutPictureBG{
  \AtPageLowerLeft{%
    \raisebox{2\baselineskip}{\makebox[\paperwidth]{\begin{minipage}{19cm}
    \footnotesize \textbf{NOTE:} This paper has been accepted for publication at IEEE/ACM Transactions on Networking 2023 (DOI: \href{https://doi.org/10.1109/TNET.2023.3269983}{10.1109/TNET.2023.3269983}). ©2023 IEEE. Personal use of this material is permitted. Permission from IEEE must be obtained for all other uses.
    \end{minipage}}}%
  }
}

\maketitle

\IEEEdisplaynontitleabstractindextext


\section{Introduction}

\IEEEPARstart{N}{etwork} modeling is arguably one of the key tools when designing, building, and evaluating computer networks, even since the early days of networking~\cite{kleinrock1970analytic}. Network models are used in protocol design, performance evaluation, or network planning, just to cite a few examples. The two most widespread network modeling techniques are analytical models based on Queuing Theory (QT), and packet-level \mbox{simulators~\cite{robertazzi2000computer, fujimoto2003large}}.

However, the evolution of computer networks, especially concerning complexity and traffic characteristics, highlights some of the limitations of classical modeling techniques. Despite their tremendous success and widespread usage, some scenarios require more advanced techniques capable of accurately modeling complex traffic characteristics, while scaling to large real-world networks. 

Especially, two relevant applications can benefit from advanced network modeling techniques: Network Digital Twins (NDT)~\cite{almasan2022digital}, and network optimization tools. Commonly, an NDT is referred to as a virtual replica of a physical network that can accurately mimic its behavior and can make performance predictions for any given input condition (e.g., traffic, topology change, or new routing configuration). In other words, an NDT is an accurate network model that can support a wide range of network configurations and that can accurately model the complex non-linear behaviors behind real-world networks. As a result, NDTs can be used to produce accurate performance predictions, carry out what-if analysis, or perform network optimization by pairing it with an optimization algorithm~\cite{nguyen2021digital, almasan2022digital}.

In the context of network optimization, \emph{we can only optimize what we can model}. Optimization algorithms operate by searching the network configuration space (e.g., to find an alternative routing scheme). For each configuration, a network model is used to estimate the resulting performance to see if it fulfills the optimization goal (e.g., minimize delay~\cite{mirjalili2019genetic}). To achieve efficient online optimization, it is essential an accurate and fast network model.

State-of-the-art modeling techniques have important limitations in effectively supporting the stringent requirements of current packet-switched networks. Queuing Theory imposes strong assumptions on the packet arrival process (Poisson traffic generation), which often is not sufficient to model real-world networks~\cite{xu2018experience}. Internet traffic has been extensively analyzed in the past two decades~\cite{ArfeenRoleWeibull2019, KaragiannisNonStationaruy2004, KaragiannisLongRange2004, KreschPoissonBased2011, PaxsonWanTraffic1995} and, despite the community has not agreed on a universal model, there is consensus that in general aggregated traffic shows strong autocorrelation and a heavy-tail~\cite{popoola2017empirical}.

Alternatively, packet-level simulators can accurately model networks. However, this comes at a  high computational cost. The cost of a simulator depends linearly on the number of packets forwarded, which can be in the range of millions per second on a single 1Gbps link. In consequence, they are slow and impractical when considering large networks with realistic traffic volumes. This also severely limits its applicability to online network optimization, given the hard time constraints of such types of applications.

In this context, Deep Learning (DL) offers an extraordinary set of techniques to build accurate data-driven network models. DL models can be trained with real data, without making any assumptions about physical networks. This enables building models with unprecedented accuracy by modeling the entire range of non-linear and multidimensional characteristics.

In this paper, we first make a systematic analysis of the performance of DL techniques for network modeling, using classical discrete-event network simulators as a baseline. Specifically, we analyze the performance of Multilayer Perceptron-based (MLP), Recurrent Neural Network-based (RNN), and Graph Neural Network-based (GNN) models. We find that classical DL techniques, such as MLPs and RNNs, are not practical enough for network modeling as they fail to provide accurate estimates when the network scenario differs from the examples seen during training (e.g., link failure). More recently, GNNs have been proposed as a novel neural network architecture specifically designed to learn over graph-structured data. They have been successfully used in other domains, such as quantum chemistry~\cite{gilmer2017neural} or logistics~\cite{kosasih2021machine}. However, in our analysis, we find that standard GNNs~\cite{gilmer2017neural} do not work well for network modeling and that we need a custom GNN architecture to model computer networks.

\begin{figure}[!t]
\centering
\includegraphics[width=\columnwidth]{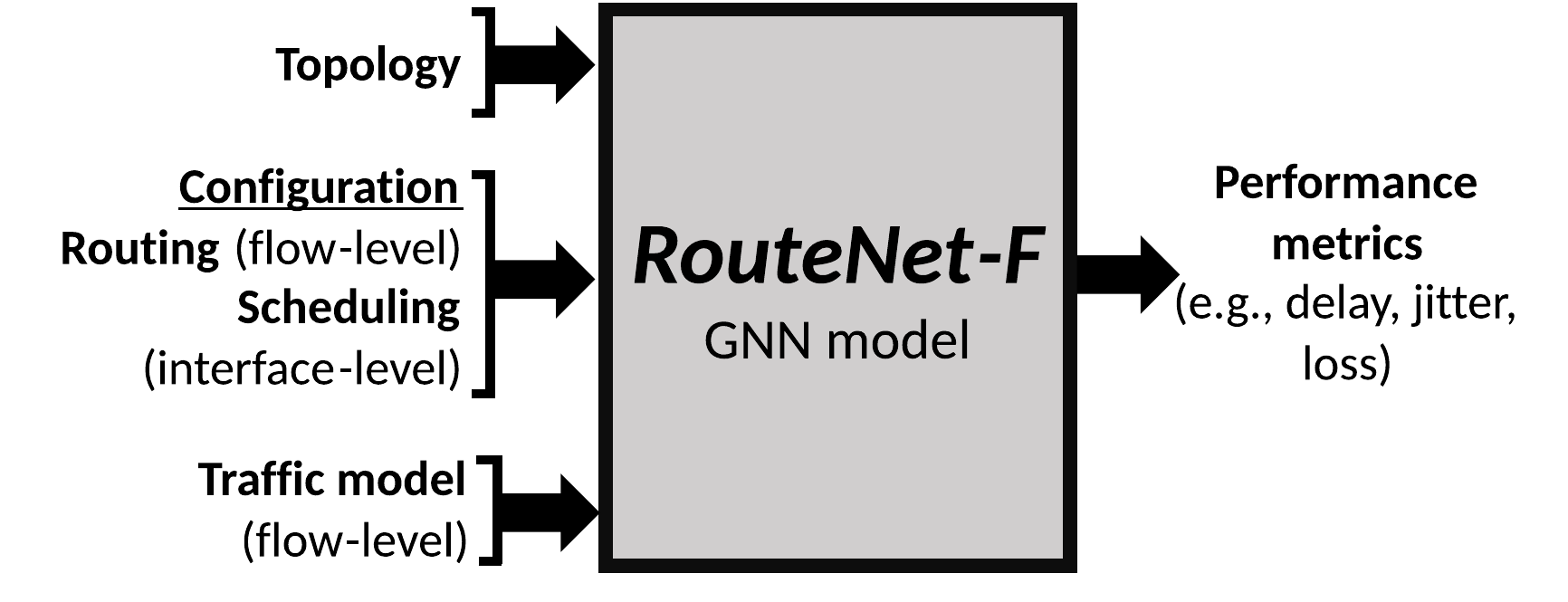}
\caption{Black-box representation of RouteNet-F.}
\vspace{-0,5cm}
\label{fig:RouteNet-F-black-box}
\end{figure}

As a result, we propose RouteNet-Fermi (RouteNet-F), a GNN architecture for network modeling. RouteNet-F shares the same goals as Queuing Theory. It provides performance estimates (delay, jitter, and packet loss) on given network scenarios (Figure~\ref{fig:RouteNet-F-black-box}) with remarkable accuracy and speed. The proposed model is not limited to Markovian traffic as Queuing Theory; it supports arbitrary models (including autocorrelated processes) which better represent the properties of real-world traffic~\cite{popoola2017empirical}. Interestingly, it also overcomes one of the main limitations of DL-based models: RouteNet-F generalizes and provides accurate estimates in network scenarios not seen in training (e.g., different topologies, traffic matrices, routing configurations). We benchmark RouteNet-F against a state-of-the-art DL-based model (MimicNet~\cite{zhang2021mimicnet}) and with a state-of-the-art queuing theory model. We show that our model outperforms both baselines in all the scenarios, achieving a 5.64\% error when tested in a dataset with packet traces coming from a real-world network, an 11\% error when evaluated in a physical testbed, and a 6.24\% error when estimating the delay on networks over a large dataset with 1,000 network samples, with topologies ranging from 50 to 300 nodes.

As any Deep Learning model~\cite{zhang2021mimicnet, yangDeepQueueNet2022, rao2018deep}, RouteNet-Fermi does not provide strong mathematical performance guarantees. However, the error of the estimates produced by the model is strongly bounded. The minimum estimated delay assumes no queuing across the path while the maximum assumes that all the queues are full. RouteNet-Fermi will not produce delay estimates outside these bounds.

The implementation of the model used in the evaluation of this paper is publicly available at~\cite{github-fermi}.

\section{Challenges of data-driven Network Modeling}\label{sec:challenges}

This section describes the main challenges that data-driven solutions need to address for network modeling. These challenges drove the core design of RouteNet-Fermi.

\textbf{Traffic models:} 
Networks carry different types of traffic, so, supporting arbitrary stochastic traffic models is crucial. Experimental observations show that traffic on the Internet has strong autocorrelation and heavy-tails~\cite{popoola2017empirical}. In this context, it is well-known that the main limitation of Queuing Theory is that it fails to provide accurate estimates on realistic Markovian models with continuous state space, or non-Markovian traffic models. The challenge for DL-based modeling is: How can we design a neural network architecture that can accurately model realistic traffic models?

\textbf{Training and Generalization:} One of the main differences between analytical modeling (e.g., QT) and data-driven modeling is that the latter requires training. In DL, training involves obtaining a \emph{representative} dataset of network measurements. The dataset needs to include a broad spectrum of network operational regimes, ranging from different congestion levels to various routing configurations, among others. In other words, the DL model can predict only scenarios it has previously seen. Note that this is a common property of all neural network architectures.

Ideally, we would obtain this training dataset from a production network, since they commonly have systems in place to measure performance. However, it would be difficult to obtain a \emph{representative} dataset. As we mentioned previously, we would need to measure the production network when it is experiencing extreme performance degradation as the result of link failures, incorrect configurations, severe congestion, etc. However, these situations are not common in production networks, which limits the ability to generate the training dataset. A reasonable alternative is creating these datasets in controlled testbeds, where it is possible to use different traffic models, implement a broad set of configurations, and replicate a wide range of network failures. Thus, the DL model can be trained on samples from this testbed and then, applied to production networks. Hence, the research challenge is: how to design a DL model that can provide accurate estimates in networks not seen during training? This includes topologies, traffic, and configurations (e.g., queue scheduling, routing) different from those seen in the training network testbed.

Leveraging a testbed that is smaller than a production network creates another challenge: the generalization to \emph{larger} networks. Real-world networks include hundreds or thousands of nodes, and building a network testbed at this scale is typically unfeasible. As a result, the DL model should be able to learn from datasets with samples of small network testbeds and predict metrics for considerably larger networks, e.g., by a factor of 10-100x. Generalizing to larger networks, or graphs in general, is currently an open research challenge in the field of GNNs~\cite{ma2021subgroup, yehudai2021local}. 

\textbf{Quality of Service and Scheduling policies:} 
A key requirement of modern networks is supporting Quality of Service (QoS), usually implemented via scheduling policies and mappings of traffic flows to QoS classes. Hence, a DL model should be able to predict the performance of the input traffic flows with their associated QoS class, similarly to how QT models support a wide range of scheduling policies~\cite{al2009performance, kapadia1984analysis}.

\section{Limitations of current Network Modeling Techniques} \label{sec:limitations}

This section explores the performance of different DL models with respect to an accurate packet-level simulator and discusses the main limitations of existing network modeling techniques.

\subsection{Simulation as a Network Modeling Technique}\label{subsec:simulators}
Network simulators reproduce the network behavior at the granularity of packet events~\cite{fujimoto2003large,weingartner2009performance}. This way, they can offer excellent accuracy and can be easily extended to include virtually any feature, such as packet scheduling, wireless networks, etc. Some simulators, such as OMNET++~\cite{varga2001discrete} or ns3~\cite{riley2010ns}, are widely used and maintained. 

However, their main limitation is the simulation time, especially for networks with high-speed links (10Gbps and above). Hence, depending on the amount of traffic found in the target network, it may become unfeasible to simulate the network~\cite{zhang2021mimicnet}.

To illustrate this limitation, we simulate different topologies using the OMNET++ simulator to calculate the delay of a set of source-destination flows [CPU Intel Xeon Silver 4210R @ 2.40 GHz]. Network topologies are artificially generated using the Power-Law Out-Degree Algorithm from~\cite{palmer2000generating} and a traffic distribution that follows a Poisson process.

\begin{figure}[!tbp]
\centering
\includegraphics[width=0.95\columnwidth]{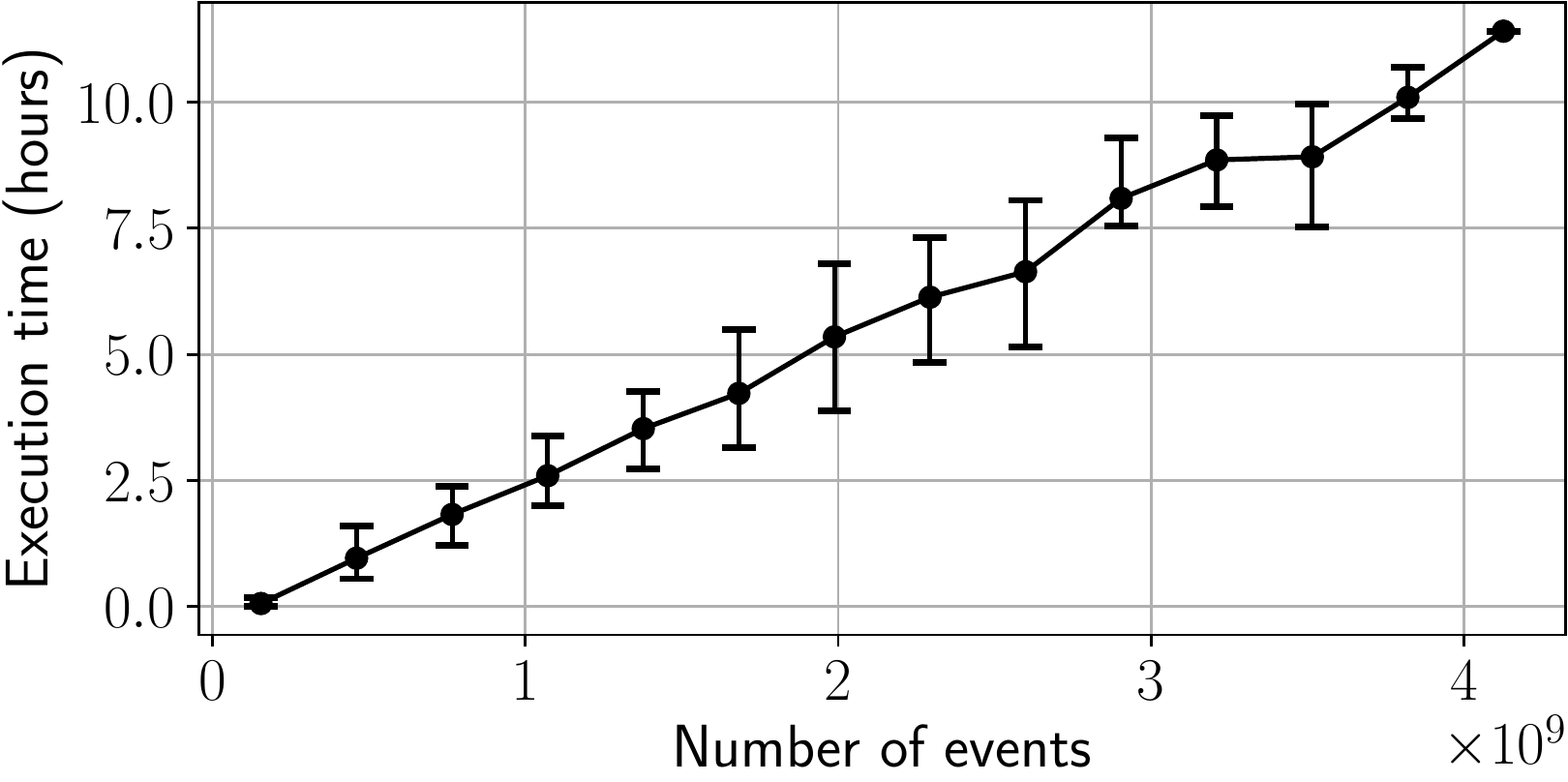}
\caption{Simulation time depending on the number of processed events.}
\label{fig:sim:events}
\end{figure}

Figure~\ref{fig:sim:events} shows the simulation time of such networks depending on the number of events. Here, an event refers to a transition in the status of the network (e.g., adding a new packet to a queue). We can see that the simulation time increases linearly and that simulating 4 billion events takes more than 11 hours. Although 4 billion events may appear a large figure, consider that a 10 Gbps link transmitting regular Ethernet frames translates to $\approx$820k events per second or 247 million events in 5 minutes of network activity for a single link. For example,  in our experiments, the simulator takes around 8h to compute the performance metrics of a 300-node network.

So, the main limitation of packet-level simulators is the simulation time. On the contrary, packet simulators offer unrivaled accuracy and can simulate virtually any scenario, from different routing configurations to replaying packet traces to simulate unknown traffic models. Because of this, hereafter, we consider the results from the simulator as the ground truth for the evaluations in this paper.

\subsection{Neural Networks as Network Modeling Techniques}

The following sections review the performance of three common Neural Network (NN) architectures in the order of increasing complexity. First, we evaluate the Multilayer Perceptron, one of the simplest NNs. Next, Recurrent Neural Networks which are designed to work with sequences. Finally, we directly input the network into a Graph Neural Network specifically designed to work with graphs. The objective is to create a network model with the NN that can predict performance parameters for input networks with a wide range of characteristics. We are especially interested in the following parameters:
\begin{itemize}
    \item \textbf{Accuracy:} How close is the prediction to simulation values?
    \item \textbf{Different Routing:} Does the accuracy degrade if we change the routing configuration?
    \item \textbf{Link failures:} Quantify  if link failures affect the quality of predictions. 
\end{itemize}

We train and test the three neural networks with the same dataset, obtained from simulations with  OMNET++. The input values are the network characteristics (topology, routing configuration, traffic model and intensity, etc), and the output values are the delay for each path. Hence, all the errors are computed with respect to the values of the simulator. We use four different datasets:

\begin{itemize}
    \item \textbf{Traffic Models:} In it, we consider traffic models that are non-Poisson, auto-correlated, and with heavy tails. Table~\ref{tab:sim_variables} details the different traffic models.
    \item \textbf{Same Routing:} Where the testing and training datasets contain networks with the \textit{same} routing configurations.
    \item \textbf{Different Routing:} Where the training and testing datasets contain networks with \textit{different} routing configurations.  
    \item \textbf{Link failures:} Here, we iteratively remove one link of the topology to replicate a link failure, until we transform the network graph into a connected acyclic graph. This scenario is the most complex since a link failure triggers a change both in the routing and the topology.
\end{itemize}

To compare the different techniques, we compute the prediction error with respect to the accurate performance values produced by the simulator. Particularly we use the following error metrics: $(i)$~Mean Absolute Percentage Error (MAPE), $(ii)$~Mean Squared Error (MSE),  $(iii)$~Mean Absolute Error (MAE), and $(iv)$~Coefficient of Determination (R\textsuperscript{2}).

\subsection{Multilayer Perceptron}\label{sec:sec:mlp}

A Multilayer Perceptron (MLP) is a basic kind of NN from the family of feedforward NNs~\cite{pal1992multilayer}. In short, input data is propagated unidirectionally from the input neuron layer to the output layer. There may be an arbitrary number of hidden layers between these two layers, and this determines how deep is the NN.

\subsubsection{Design}

Several works have leveraged an MLP to predict network performance metrics~\cite{sadeghzadeh2008mlp, mestres2018understanding, wang2017machine}. Based on this work, we have built an MLP to predict the mean delay for each source-destination pair of nodes of a given network. The MLP has 8,280 inputs and two hidden layers with 4096 neurons and uses Rectified Linear Units (ReLU) as activation functions.

\subsubsection{Evaluation}

Table~\ref{tab:traffic_model_mlp_rnn} presents the error when predicting the network delay with respect to the results produced by the network simulator, including several traffic models. We can see that the MLP offers good accuracy for Poisson traffic, but the error increases significantly for the rest of the traffic models showing a MAPE between 23\% and 84\%.

\begin{table}[!t]
\caption{Delay prediction using an MLP and an RNN for different traffic models. The error is computed w.r.t. simulation results.}
\label{tab:traffic_model_mlp_rnn}
\centering
\resizebox{\columnwidth}{!}{%
\begin{tabular}{ccccccccc}
\toprule
     & \multicolumn{4}{c}{MLP}                 & \multicolumn{4}{c}{RNN}                \\
     \cmidrule(lr){2-5}
 \cmidrule(lr){6-9}
     & MAPE & MSE  & MAE  & R\textsuperscript{2} & MAPE & MSE  & MAE  & R\textsuperscript{2} \\
     \midrule
Poisson & 12.3\% & 0.103 & 0.122 & 0.801 & 10.0\% & 0.071 & 0.084 & 0.862 \\
Deterministic & 23.9\% & 0.309 & 0.160 & 0.044 & 13.1\% & 0.083 & 0.070 & 0.743 \\
On-Off & 30.4\% & 0.438 & 0.240 & 0.002 & 15.2\% & 0.065 & 0.082 & 0.851 \\
A. Exponentials & 84.5\% & 1.013 & 0.308 & -1.935 & 14.0\% & 0.070 & 0.072 & 0.7961 \\
M. Exponentials & 57.1\% & 1.058 & 0.363 & -1.679 & 57.8\% & 0.528 & 0.457 & -0.338 \\
Mixed & 41.2\% & 0.351 & 0.269 & 0.052 & 17.5\% & 0.036 & 0.080 & 0.900 \\
\bottomrule
\end{tabular}%
}
\end{table}

Likewise, Table~\ref{tab:mlp_rnn_routing_failures} shows the error of predicting the delay for the datasets with the same/different routing and link failures. We can see that the MLP cannot offer an accurate estimate when predicting the delay of a previously unseen routing configuration (1150\% of error). This is due to the internal architecture of the MLP. During training, the MLP performs overfitting, meaning that the model \emph{only} learns about the initial network topology used for training and not for any others. When we input a new topology, it does not have sufficient information to make an accurate prediction.

\begin{table}[!b]
\caption{Delay prediction using an MLP and an RNN for the same and different routing configurations w.r.t. those seen during training, and considering various link failures. The error is relative to simulation results.}
\label{tab:mlp_rnn_routing_failures}
\centering
\resizebox{\columnwidth}{!}{%
\begin{tabular}{ccccccccc}
\toprule
     & \multicolumn{4}{c}{MLP}                 & \multicolumn{4}{c}{RNN}                \\
     \cmidrule(lr){2-5}
 \cmidrule(lr){6-9}
     & MAPE & MSE  & MAE  & R\textsuperscript{2} & MAPE & MSE  & MAE  & R\textsuperscript{2} \\
     \midrule
Same Routing & 12.3\% & 0.103 & 0.122 & 0.801 & 10.0\% & 0.071 & 0.084 & 0.862 \\
Diff. Routing & 1150\% & 28.3 & 2.96 & -40.0 & 30.5\% & 0.553 & 0.282 & 0.197 \\
Link Failures & 125\% & 3.69 & 1.03 & -0.191 & 63.8\% & 2.971 & 0.870 & 0.0417 \\
\bottomrule
\end{tabular}%
}
\end{table}

\begin{figure}[!t]
\centering
\includegraphics[width=0.65\columnwidth]{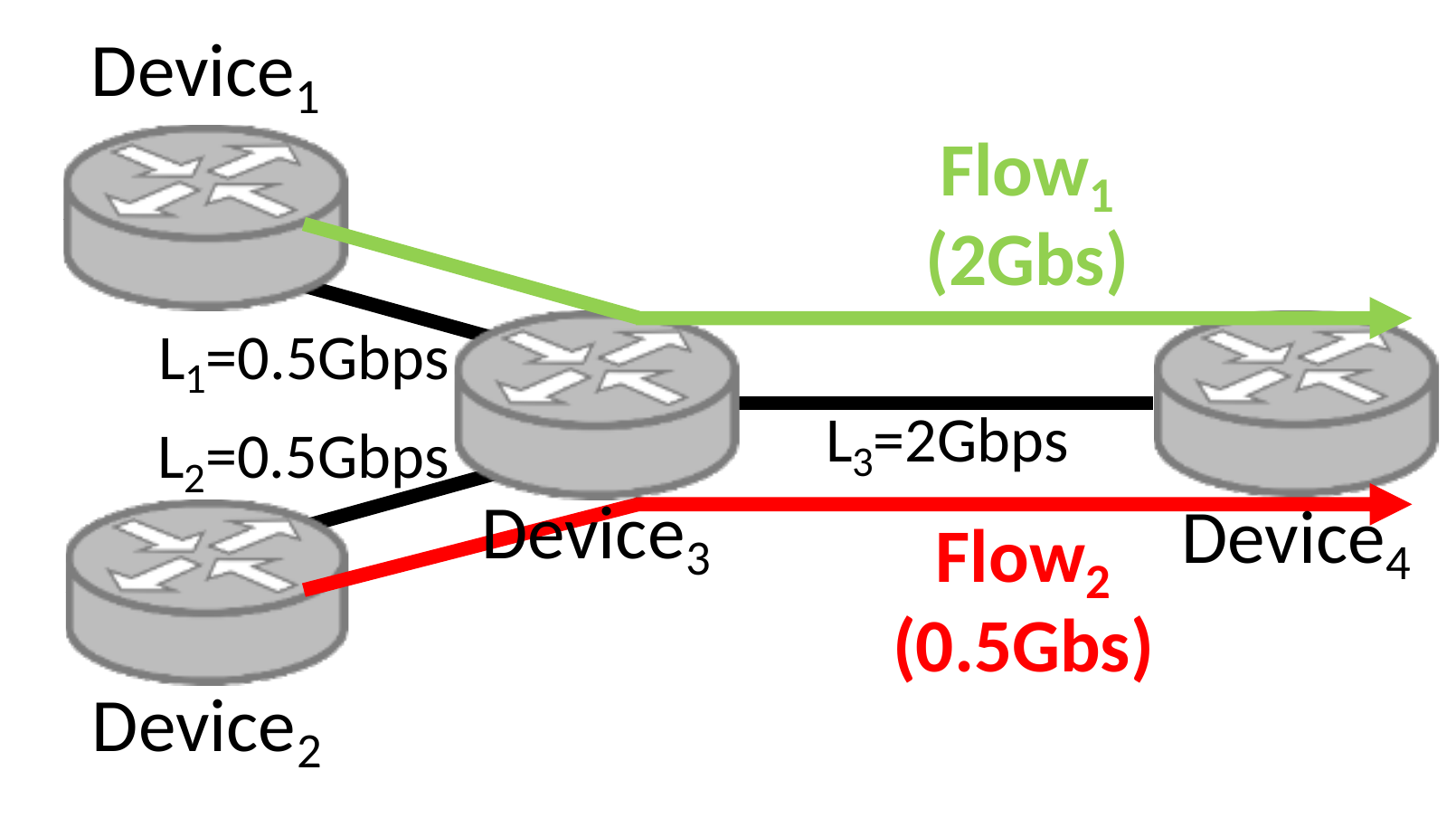}
\caption{Sample Topology with 4 nodes, three links, and two flows.}
\label{fig:sample_topology}
\end{figure}

\subsection{Recurrent Neural Networks}
Recurrent Neural Networks (RNN) are a more advanced type of NN. They have shown excellent performance when processing sequential data~\cite{mikolov2011extensions}. This is mainly because they connect some layers to the previous ones, which gives them the ability to keep the state along sequences.

\subsubsection{Design}

\begin{figure}[!t]
\centering
\includegraphics[width=0.65\columnwidth]{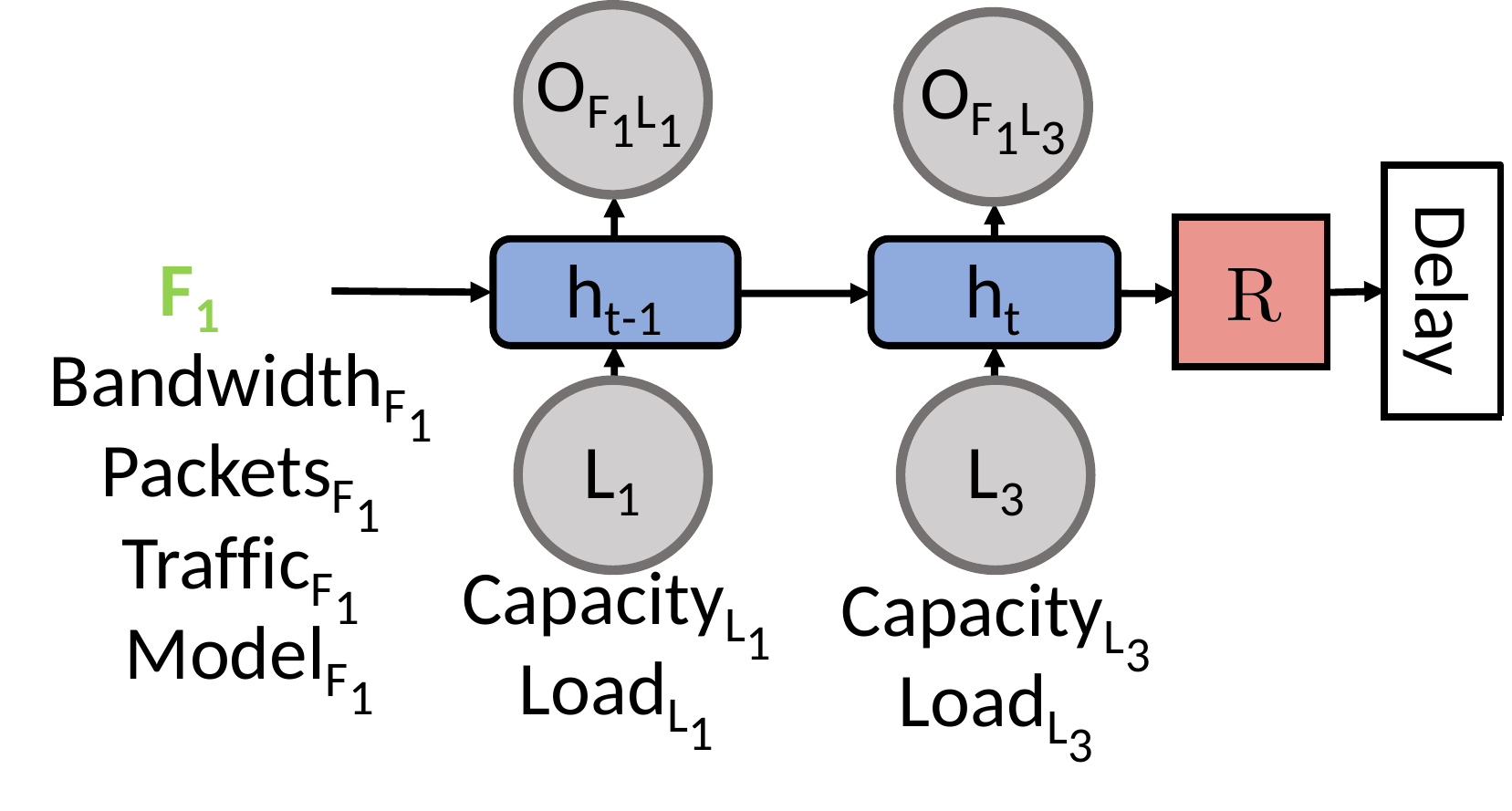}
\caption{Recurrent Neural Network model for the Sample Topology (Figure~\ref{fig:sample_topology})} 
\label{fig:rnn_model}
\end{figure}

Several works~\cite{belhaj2009modeling, mohammed2019machine, naseer2018enhanced} propose RNNs as a way to predict network performance. In this experiment, we build a sequential model with an RNN (Figure~\ref{fig:rnn_model}). Particularly, we choose a Gated Recurrent Unit (GRU). 

We initialize the state of each path with the sequence of nodes in the path and the features of the traffic model (e.g., packets, bandwidth, $\lambda$, $\epsilon$, $\alpha$, or on-off time), and we update the state of each link across the path. As an example, Figure~\ref{fig:rnn_model} shows the structure of an RNN to model the sample network from Figure~\ref{fig:sample_topology}. We can see that the path of $flow_1$ is composed of $L_1$ and $L_3$. Once the path state has been computed, an MLP with 2 hidden layers computes the final output.

\subsubsection{Evaluation}

We train the RNN with the same datasets as the previous subsection. Although the RNN supports better different traffic models than the MLP (Table~\ref{tab:traffic_model_mlp_rnn}), it still struggles to produce accurate predictions when there are routing or topology changes (Table~\ref{tab:mlp_rnn_routing_failures}), especially for different routing configurations (30\% error), or when removing links (63\%).

The reason behind the lack of capability of RNNs to understand routing changes and link failures is due to its internal architecture. RNNs can accommodate different end-to-end paths in the network (i.e., series of routers and links), thereby, making it easier to perform predictions for paths never seen in the training phase. However, this structure cannot store and update the status of individual links in the topology due to the inter-dependency between links and traffic flows (i.e., routing). In other words, if the status of a link changes, it affects several flows, and vice-versa. This generates circular dependencies that RNNs are not able to model (see more details in Sec.~\ref{sec:routenet_fermi}).

\subsection{Graph Neural Networks}

Networks are fundamentally represented as graphs, where networking devices are the graph nodes and the links connecting devices are the graph edges. This interconnection translates to the fact that the different elements in the network are dependent on each other. Since most standard DL models (e.g., MLP, RNN) assume independent flow-level data points. This renders them inaccurate for our use case. Hence, a model that is capable of processing directly the network graph is arguably more desirable for network modeling, because it will be able, not only to obtain information from the individual nodes and edges but also from the underlying data structure (i.e., the relationships between the different elements)~\cite{suarez2021graph}.

GNNs~\cite{scarselli2008graph} are a type of neural network designed to work with graph-structured data. They have two key characteristics that make them a good candidate for a network model. First, GNNs have the ability to store node-level hidden states and update them in each iteration. Second, they process the input data directly as a graph, both during training and inference. This means that the internal structure of the neural network depends on the input graph. Hence, they can dynamically adapt to the underlying dependencies between the different elements of a network~\cite{battaglia2018relational,zhou2018graph}. The latter is of special importance, since changes in the network graph (e.g., routing modifications, link failures) are the main limitation of other DL-based models, such as MLPs and RNNs, as we have seen in the previous subsections.

In this section, we build a standard GNN model to predict the mean per-flow delay in networks.

\subsubsection{Design}

We implement a Message-Passing Neural Network (MPNN), a powerful state-of-the-art GNN architecture that can efficiently capture dependencies between the elements of input graphs~\cite{gilmer2017neural}. We define a graph $G$ described as a set of vertices (or nodes) $E$ and a set of edges (or links) $V$. Each node has a set of features $x_v$, and edges also have some features $e_{vw}$. The execution of an MPNN can be divided into three phases, an initialization phase, a message-passing phase, and a readout phase. The first one defines a hidden state ($h_v^0$) using the node features ($x_v$). The second one is an iterative process that runs for $T$ time steps and that is defined by two functions: the message function $M_t$ and the update function $U_t$. During this phase, node hidden states ($h_v^t$), represented as fixed-size vectors, attempt to encode some \textit{meaningful} information, and are iteratively updated by exchanging messages $m_v^{t+1}$ with their neighbors:

\begin{align}
m_v^{t+1} = \sum_{w \in N(v)}{M_t(h_v^t,h_w^t,e_{vw})} \label{eq:message}\\
h_v^{t+1} = U_t(h_v^t,m_v^{t+1})
\label{eq:update}
\end{align}

where $N(v)$ represents the neighbors of $v$ in graph $G$. Finally, the readout phase computes the output vector using a readout function $R$ that takes as input the final hidden states~$h_v^T$:

\begin{equation} \label{eq:readout}
\hat{y} = R(\{h_v^T | v \in G\})
\end{equation}

In our case, the input of the MPNN is the network topology graph. It performs $T$$=$$4$ message-passing iterations and the hidden state dimension is 32. The readout function is implemented with a two-layer fully connected NN with ReLUs as activation functions for the hidden layers and a linear activation for the output layer.

\subsubsection{Evaluation}

We evaluate the accuracy of the MPNN model when predicting the mean flow-level delay, as in previous subsections. Table~\ref{tab:mpnn_gnn_routing_failures} presents the delay prediction errors in the same scenarios of the previous experiments: routing configurations, both seen and not seen during training, and link failures. Unfortunately, the results are similar to those of the RNN: the routing configurations from the training dataset are easy to predict, with an error as low as  3\%, while new routing configurations and link failures increase the error significantly (respectively, 50\% and 125\% of MAPE), thus showing even larger errors than the RNN model.

\begin{table}[!b]
\caption{Delay prediction using an MPNN for the same and different routing configurations w.r.t. those seen during training, and considering various link failures. The error is relative to simulation results.}
\label{tab:mpnn_gnn_routing_failures}
\centering
\resizebox{0.7\columnwidth}{!}{%
\begin{tabular}{ccccc}
\toprule
     & MAPE & MSE  & MAE  & R\textsuperscript{2} \\
     \midrule
Same Routing & 3.0\% & 0.002 & 0.016 & 0.994 \\
Diff. Routing & 50.0\% & 0.609 & 0.307 & 0.115 \\
Link Failures & 82.2\% & 3.41 & 0.949 &  -0.099  \\
\bottomrule
\end{tabular}%
}
\end{table}

The main reason behind the poor accuracy of the MPNN model is that the architecture of this model is directly built based on the network topology without taking into account the paths traversed by different traffic flows (i.e., the routing configuration), which is a fundamental property to understand inter-dependencies between flows and links. More specifically, when we test the model with the same routing configuration, it learns the relationships between flows and links. However, when we change this configuration, those previously-learned relationships are no longer valid, and the model is not able to capture the relationships between elements in the new scenario.

This intuition is better understood with an extreme packet loss example. Let’s suppose we have the sample network in Figure~\ref{fig:sample_topology}. The first flow (Flow\textsubscript{1}) is transmitting at a rate of 2Gbps, and the second one (Flow\textsubscript{2}) is transmitting at a rate of 0.5Gbps. As we can see, L\textsubscript{1} has a maximum capacity of 0.5Gbps. Because of this, Flow\textsubscript{1} will experience a large packet loss, which causes the traffic of Flow\textsubscript{1} at L\textsubscript{3} to be at most 0.5Gbps. Hence, instead of having 2Gbps+0.5Gbps of aggregated traffic at L\textsubscript{3}, it will only have 1Gbps. Now, Flow\textsubscript{2}, which initially could have experienced a lot of network congestion when going through the 2Gbps link will experience none as the state of the link changed.

Knowing this, we can see how there is a circular dependency between the flows and links found in the network. At the same time, the state of a flow depends on the state of the links they traverse, and the state of the links depends on the state of the flows passing through them.

If we apply a standard GNN over this example, the state of each flow is not updated at each hop. Therefore, the GNN does not have a structure that represents how the delay depends on both the links (topology) and the flows that go through each specific router.

Hence, we conclude that feeding the MPNN directly with the network topology graph is not sufficient to accurately perform network modeling. However, more complex and customized GNN-based architectures can still be powerful for modeling the inter-dependencies between the different network elements and generalizing over new network scenarios by exploiting the underlying graph structure.

\section{RouteNet-Fermi}\label{sec:routenet_fermi}

\begin{figure*}
    \centering
    \includegraphics[width=0.95\textwidth]{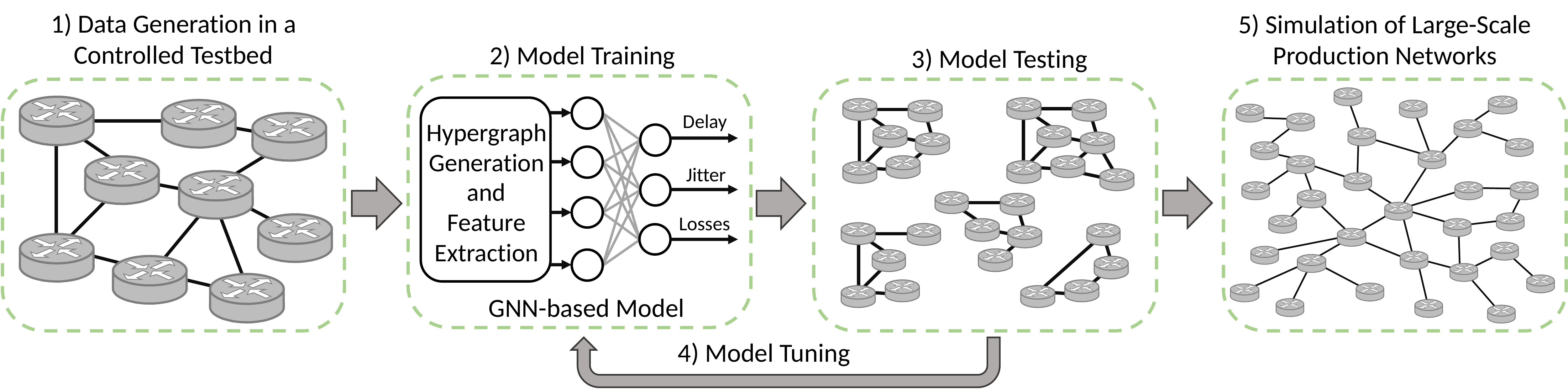}
    \caption{End-to-end workflow of RouteNet-F. 1) Collection of small-scale observations coming from a controlled environment, 2) Model training, 3) Model testing with various configurations (e.g., routing, scheduling) never seen during the training phase, 4) hyper-parameter tunning to balance highly accurate predictions with performance, 5) Simulation of large-scale production networks. One of the main advantages of RouteNet-F is that usually time-consuming steps like 1), 2), 3), and 4) are all done at small scales and, therefore, are fast as well.}
    \label{fig:workflow}
\end{figure*}

This section describes the internal GNN architecture of RouteNet-Fermi (hereafter referred to as RouteNet-F). This GNN-based model implements a custom three-stage message-passing algorithm that represents key elements for network modeling (e.g., topology, queues, traffic flows). \mbox{RouteNet-F} supports a wide variety of features present in real-world networks, such as multi-queue QoS scheduling policies or complex traffic models.

Figure~\ref{fig:RouteNet-F-black-box} shows a black-box representation of \mbox{RouteNet-F}. The input of this model is a network sample, defined by: a network topology, a routing scheme (flow level), a queuing configuration (interface level), and a set of traffic flows characterized by some parameters. As output, the model produces estimates of relevant performance metrics at a flow-level granularity (e.g., delay, jitter, packet loss). Figure~\ref{fig:workflow} shows the end-to-end workflow to train, validate, and use RouteNet-Fermi.

\subsection{Model Description}\label{subsec:model}
RouteNet-F is based on two main design principles: \mbox{$(i)$~\textit{finding a good representation}} of the network components supported by the model (e.g., traffic models, routing, queue scheduling), and $(ii)$~\textit{exploit scale-independent features} of networks to accurately scale to larger networks unseen during training. These two aspects are further discussed in the next two subsections.

\subsection{Representing network components and their relationships}

First, let us define a network as a set of source-destination flows $\mathcal{F}\!=\!\{f_i:i\!\in\!(1,...,n_f)\}$, a set of queues on $\mathcal{Q}\!=\!\{q_j: j \in (1,...,n_q)\}$, and a set of links \mbox{$\mathcal{L} = \{l_k: k\!\in\!(1,...,n_l)\}$}. According to the routing configuration, flows follow a source-destination path. Hence, we define flows as sequences of tuples with the queues and links they traverse \mbox{$f_i\!=\!\{(q^{i,1},l^{i,1}),...,(q^{i,M},l^{i,M})\}$}, where $M$ is the path length of the flow (number of links). Let us also define $Q_f(q_j)$ and $L_f(l_k)$ as functions that respectively return all the flows passing through a queue $q_j$ or a link $l_k$. Also, $L_q(l_k)$ is defined as a function that returns the queues $q_{l_k}\in \mathcal{Q}$ injecting traffic into link $l_k$ (i.e., the queues at the output port to which the link is connected).

Following the previous notation, RouteNet-F considers an input graph with three main components: $(i)$ the physical links~$\mathcal{L}$ that shape the network topology, $(ii)$ the queues $\mathcal{Q}$ at each output port of network devices, and $(iii)$ the active flows $\mathcal{F}$ in the network, which follow some specific src-dst path (i.e., sequences of queues and links). Traffic in flows is generated from a given traffic model. From this, we can extract three basic principles:

\begin{enumerate}
\item The state of flows (e.g., delay, throughput, packet loss) is affected by the state of the queues and links they traverse (e.g., queue/link utilization).
\item The state of queues (e.g., occupation) depends on the state of the flows passing through them (e.g., traffic volume, burstiness).
\item The state of links (e.g., utilization) depends on the states of the queues that can potentially inject traffic into the link, and the queue scheduling policy applied over these queues (e.g., Strict Priority, Weighted Fair Queuing).
\end{enumerate}

Formally, these principles can be formulated as follows:
\begin{align}
\boldsymbol{h}_{f_i}\!=\!G_f(\boldsymbol{h}_{q^{i,1}},\boldsymbol{h}_{l^{i,1}},...,\boldsymbol{h}_{q^{i,M}},\boldsymbol{h}_{l^{i,M}})\label{eq:g_f}\\
\boldsymbol{h}_{q_j} = G_q(\boldsymbol{h}_{f_1},...,\boldsymbol{h}_{f_I}), \quad f_i \in Q_f(q_j) \label{eq:g_q}\\
\boldsymbol{h}_{l_k} = G_l(\boldsymbol{h}_{q_1},...,\boldsymbol{h}_{q_J}), \quad q_j \in L_q(l_j) \label{eq:g_l}
\end{align}

Where $G_f$, $G_q$, and $G_l$ are some unknown functions, and $\boldsymbol{h}_f$, $\boldsymbol{h}_q$ and $\boldsymbol{h}_l$ are latent variables that encode information about the state of flows $\mathcal{F}$, queues $\mathcal{Q}$, and links $\mathcal{L}$ respectively. Note that these principles define a circular dependency between the three network components ($\mathcal{F}$, $\mathcal{Q}$, and $\mathcal{L}$) that must be solved to find latent representations satisfying the equations above.

\begin{figure}[!t]
\centering
\includegraphics[width=0.99\columnwidth]{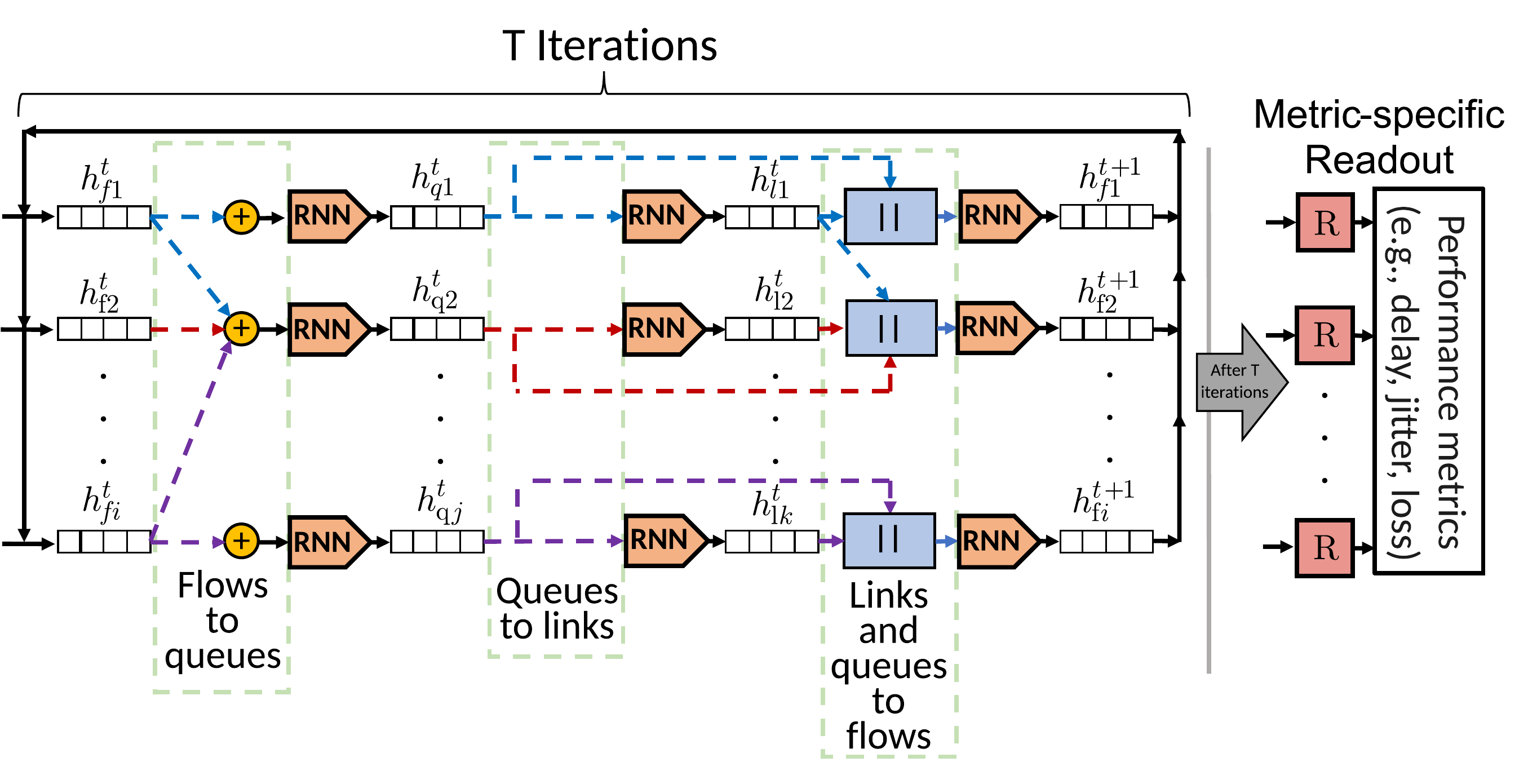}
\caption{Schematic representation of RouteNet-F.} 
\label{fig:architecture}
\end{figure}

To solve the circular dependencies defined in Equations~\mbox{\eqref{eq:g_f}-\eqref{eq:g_l}}, RouteNet-F implements a three-stage message passing algorithm that combines the states of flows $\mathcal{F}$, queues $\mathcal{Q}$, and links $\mathcal{L}$, and updates them iteratively. Finally, it combines these states to estimate flow-level delays, jitters, and packet loss. Figure~\ref{fig:architecture} shows a schematic representation of the internal three-stage message-passing architecture of this model.

Algorithm~\ref{alg:architecure} describes the architecture of RouteNet-F.  First, hidden states $\boldsymbol{h}_f$, $\boldsymbol{h}_q$, and $\boldsymbol{h}_l$ are initialized using the functions $HS_f$, $HS_q$, and $HS_l$ respectively (lines~\mbox{\ref{init-f}-\ref{init-l}}). These functions encode the initial features $\boldsymbol{x}_{f}$, $\boldsymbol{x}_{q}$, and $\boldsymbol{x}_{l}$ into fixed-size vectors that represent feature embeddings. The initial features of flows $\boldsymbol{x}_f$ are defined as an n-element vector that characterizes the flow's traffic. For example, in our case, this vector includes the average traffic volume transmitted in the flow~$\lambda$, and some specific parameters of the traffic model, such as $t_{on}$ and $t_{off}$ for On-Off traffic distributions or $\alpha$ and $\beta$ for exponential models. We set the initial features of links $\boldsymbol{x}_l$ as $(i)$ the link load~$x_{l_{load}}$, and \mbox{$(ii)$~the} scheduling policy at the output port of the link (FIFO, Strict Priority, Weighted Fair Queuing, or Deficit Round Robin). For the scheduling policy, we use a one-hot encoding. The calculation of the link load~$x_{l_{load}}$ is defined in more detail later (Sec.~\ref{subsec:scaling}). Lastly, the initial features of queues~$\boldsymbol{x}_q$ include: $(i)$ the buffer size, $(ii)$ the queue order/priority level (one-hot encoding), and $(iii)$~the weight (only for Weighted Fair Queuing or Deficit Round Robin configurations).

Once all the hidden states are initialized, the message-passing phase starts. This phase is executed for $T$ iterations (loop from line~\ref{init-loop}), where $T$ is a configurable parameter of the model. Each message passing iteration is divided into three stages, which represent respectively the message exchanges and updates of the hidden states of flows $\boldsymbol{h}_f$ (lines~\mbox{\ref{mp-path-init}-\ref{mp-path-end}}), queues $\boldsymbol{h}_q$ (lines~\mbox{\ref{mp-queue-init}-\ref{mp-queue-end}}), and links $\boldsymbol{h}_l$ (lines~\mbox{\ref{mp-link-init}-\ref{mp-link-end}}). 

Finally, the loop from line~\ref{readout} computes the different flow-level performance metrics. Here, function $R_{f_d}$ (line~\ref{q_delay}) and $R_{f_j}$ (line~\ref{r-jitter}) represent a readout function that is individually applied to the hidden states of flows as they pass through a specific link ($\boldsymbol{h}_{f,l}$). The output of these functions is the average queue occupancy and the delay variation (i.e., jitter) seen by the flow at that link. Note that different flows may experience different queue occupancies and jitter depending on their traffic properties (e.g., traffic volume, burstiness). Lastly, these link-level delay and jitter estimates are combined to compute the final flow-level delay $\hat{y}_{f_d}$ and jitter $\hat{y}_{f_j}$. This calculation is further described in Section~\ref{subsec:scaling}. Similarly, $R_{f_l}$ (line~\ref{r-loss}) is applied to the hidden states of flows $h_f$ to compute the per-flow packet loss rate.

\algnewcommand\algorithmicforeach{\textbf{for each}}
\algdef{S}[FOR]{ForEach}[1]{\algorithmicforeach\ #1\ \algorithmicdo}
\renewcommand{\algorithmicrequire}{\textbf{Input:}}
\renewcommand{\algorithmicensure}{\textbf{Output:}}
\begin{algorithm}[!t] \caption{Internal architecture of RouteNet-F.}
\begin{algorithmic}[1]
\Require{$\mathcal{F}$, $\mathcal{Q}$, $\mathcal{L}$, $\boldsymbol{x}_f$, $\boldsymbol{x}_q$, $\boldsymbol{x}_l$}
\Ensure{$\hat{y}_{f_d}$}
\ForEach {$f \in \mathcal{F}$} $\boldsymbol{h}^0_f \gets HS_f(\boldsymbol{x}_f)$ \EndFor \label{init-f}
\ForEach {$q \in \mathcal{Q}$} $\boldsymbol{h}^0_q \gets HS_q(\boldsymbol{x}_q)$ \EndFor \label{init-q}
\ForEach {$l \in \mathcal{L}$} $\boldsymbol{h}^0_l \gets HS_l(\boldsymbol{x}_l)$ \EndFor \label{init-l}
\Statex
\For{t = 0 to T-1} \label{init-loop} \Comment{\footnotesize Message Passing Phase}
    \ForEach {$f \in \mathcal{F}$} \label{mp-path-init} \Comment{\footnotesize Message Passing on Flows}
        \State $\Theta([\cdot,\cdot]) \gets FRNN(\boldsymbol{h}^t_{f},[\cdot,\cdot])$ \Comment{\footnotesize FRNN Initialization}
        \ForEach {$(q,l) \in f$}
            \State $\boldsymbol{h}^t_{f,l} \gets \Theta([\boldsymbol{h}^t_q,\boldsymbol{h}^t_l])$ \label{lin:u-flow} \Comment{\footnotesize Flow: Aggr. and Update}
            \State $\widetilde{m}^{t+1}_{f,q} \gets \boldsymbol{h}^t_{f,l} $ \Comment{\footnotesize Flow: Message Generation}
        \EndFor
    \State $\boldsymbol{h}^{t+1}_{f} \gets \boldsymbol{h}^{t}_{f,l} $
    \EndFor \label{mp-path-end}
    \ForEach {$q \in \mathcal{Q}$} \label{mp-queue-init} \Comment{\footnotesize Message Passing on Queues}
        \State $M_q^{t+1} \gets \sum_{f \in Q_f(q)}  \widetilde{m}^{t+1}_{f,q}$ \Comment{\footnotesize Queue: Aggregation}
        \vspace{0.1cm}
        \State $\boldsymbol{h}^{t+1}_q \gets U_q(\boldsymbol{h}^t_q,M_q^{t+1})$ \Comment{\footnotesize Queue: Update}
        \State $\widetilde{m}^{t+1}_{q} \gets \boldsymbol{h}^{t+1}_{q} $ \Comment{\footnotesize Queue: Message Generation}
    \EndFor \label{mp-queue-end}
    \ForEach {$l \in \mathcal{L}$} \label{mp-link-init} \Comment{\footnotesize Message Passing on Links}
    \State $\Psi(\cdot) \gets LRNN(\boldsymbol{h}^t_l,\cdot)$ \Comment{\footnotesize LRNN Initialization}
        \ForEach {$q \in L_q(l)$}
            \State ${h}^{t}_{l}  \gets \Psi(\widetilde{m}^{t+1}_{q})$  \label{lin:u-link} \Comment{\footnotesize Link: Aggr. and Update}
	    \EndFor
	    \State $\boldsymbol{h}^{t+1}_l \gets {h}^{t}_{l}$
    \EndFor \label{mp-link-end}
\EndFor \label{end-loop}
\Statex
\ForEach {$f \in F$} \label{readout} \Comment{\footnotesize Flow: Readout}
        \State $\hat{y}_{f_d} = 0$ \Comment{\footnotesize Initializing the flow delay}
        \State $\hat{y}_{f_j} = 0$ \Comment{\footnotesize Initializing the flow jitter}
        \ForEach {$(q,l) \in f$}
            \State $\hat{d}_{q} = R_{f_d}(\boldsymbol{h}^T_{f,l})/\boldsymbol{x}_{l_c}$  \label{q_delay} \Comment{\footnotesize Queuing delay}
            \State $\hat{d}_{t} = \boldsymbol{x}_{f_{ps}}/\boldsymbol{x}_{l_c}$ \label{t_delay} \Comment{\footnotesize Transmission delay}
            \State $\hat{d}_{link} = \hat{d}_{q} + \hat{d}_{t}$
            \State $\hat{y}_{f_d} = \hat{y}_{f_d} + \hat{d}_{link}$ \Comment{\footnotesize Sum of link delays along the flow} \label{lin:readout-end}
            \State $\hat{y}_{f_j} = \hat{y}_{f_j} +  R_{f_j}(\boldsymbol{h}^T_{f,l})/\boldsymbol{x}_{l_c}$ \Comment{\footnotesize Sum of link jitters along the flow} \label{r-jitter}
        \EndFor
            \State $\hat{y}_{f_l} \mathrel= R_{f_l}(\boldsymbol{h}^T_{f})$ \label{r-loss} \Comment{\footnotesize Packet loss prediction}
\EndFor
\end{algorithmic}
\label{alg:architecure}
\end{algorithm}
\setlength{\textfloatsep}{0.2cm}

\subsection{Scaling to larger networks: scale-independent features} \label{subsec:scaling}

Data-driven models typically need to see edge cases that are uncommon in real-world production networks (e.g., link failures). This means that collecting data directly from production networks requires testing configurations that might break the correct operation of the network. As a result, data-driven network models should be typically trained with data from controlled network testbeds. However, network testbeds are usually much smaller than real networks. In this context, it is essential for our model to effectively scale to larger networks than those seen during the training phase.

It is well-known that GNN models have an unprecedented capability to generalize over graph-structured data~\cite{battaglia2018relational,zhou2018graph}. In the context of scaling to larger graphs, it is also known that GNNs keep good generalization capabilities as long as the spectral properties of graphs are similar to those seen during training~\cite{ruiz2020graphon}. In our particular case, the internal message passing architecture of RouteNet-F generalizes accurately to graphs with similar structures (e.g., a similar number of queues at output ports, or a similar number of flows aggregated in queues). In practice, this means that RouteNet-F should be able to generalize to larger topologies when trained with smaller ones, as long as the networks used for training had the same spectral properties as the larger ones. More details about this can be found in Section~\ref{sec:evaluation}, which presents an empirical evaluation of RouteNet-F's generalization capabilities).

However, scaling to larger networks often entails other aspects beyond the topology size. Two key elements require special attention: link capacities and the range of output variables. First, larger networks naturally have larger link capacities. This in turn results in larger traffic aggregates on core links of the network. Such traffic intensities may fall in ranges not seen in smaller topologies. 
Second, as the network scales, it inherently has longer end-to-end paths, which result in increased end-to-end path delays. Again, some of these delays may fall outside the range of delays used when training with smaller topologies. These out-of-range parameters require devising mechanisms to effectively scale on them.

\subsubsection{Scaling to larger link capacities} \label{sec:link-cap}

In RouteNet-F (Algorithm~\ref{alg:architecure}), the most straightforward way to represent the link capacity $\boldsymbol{x}_{l_c}$ would be as an initial feature of the links' hidden states $\boldsymbol{x}_l$. However, the fact that $\boldsymbol{x}_{l_c}$ would be encoded as a numerical input value would then introduce inherent limitations to scale to larger capacity values. Indeed, scaling to out-of-distribution numerical values is recognized as a generalized limiting factor among all neural network models~\cite{engstrom2019exploring,su2019one}. 

Our approach is to exploit particularities from the network domain to find scale-independent representations for link capacities. These representations define link capacities and how they relate to other link-level features that impact performance (e.g., the aggregated traffic in the link), so they can be used to accurately estimate performance metrics (e.g., delay, jitter, packet loss). Inspired by traditional QT methods, we aim to encode in RouteNet-F the relative ratio between the arrival rates on links (based on the traffic aggregated in the link) and the service times (based on the link capacity). This enables the possibility to infer the output performance metrics of our model from scale-independent values. As a result, instead of directly using the numerical link capacity values, we introduce the \emph{link load} $x_{l_{load}}$ in the initial feature vector of links $\boldsymbol{x}_{l}$. Particularly, we compute the link load as follows:

\begin{equation} \label{eq:load}
x_{l_{load}} = \frac{1}{x_{l_c}} \sum_{f \in L_f(l_j)} \lambda_f
\end{equation}

Where $\lambda_f$ is the average traffic volume of the flows that traverse the link $l_j$, and $x_{l_c}$ is the link capacity. In other words, we compute the link load as the summation of all the traffic that would traverse the link without considering possible losses and divide it by the link capacity. Then, through the iterative message-passing process, the GNN model updates the load values after estimating the packet loss.

\subsubsection{Different output ranges}\label{subsec:out-dist}

The previous mechanism enables us to keep scale-independent features along with the message-passing phase of our model (loop lines~\mbox{\ref{init-loop}-\ref{end-loop}} in Alg.~\ref{alg:architecure}), while it is still needed to extend the scale independence to the output layer of the model. Note that in larger networks, delay values can vary with respect to those seen during the training in smaller networks. This is because flows can go through links with higher capacities, or because flows can potentially traverse longer paths. This again poses the challenge of generalizing to ranges of delays not seen in the training phase. Equivalently, this also applies to the prediction of flow jitter and packet loss.

To overcome this potential limitation, \mbox{RouteNet-F} infers delays indirectly from the mean queue occupancy on forwarding devices. Based on traditional QT models, RouteNet-F infers the flow delay as a linear combination of the estimated queuing delays (line~\ref{q_delay}) and the transmission delays after crossing a link (line~\ref{t_delay}). Note that a potential advantage with respect to traditional QT models is that the queue occupancy estimates produced by RouteNet-F can be more accurate, especially for autocorrelated and heavy-tail traffic models. 

We call the values produced by the $R_{f_d}$ function the \emph{effective queue occupancy}, which is defined as the mean queue occupancy experienced by a given flow $f_i$ as it passes through a specific forwarding device. More precisely, this value is the average number of bits that have to be served on a specific output port before the packets of flow $f_i$ are transmitted. As an example, let us consider the case of packets from a flow with low priority, which are mapped to low-priority queues. If forwarding devices implement a multi-queue Strict Priority scheduling policy, the effective queue occupancy seen by those low-priority packets should include all the bits to be served in the queues with higher priority.

The prediction of this effective queue occupancy ---~instead of directly predicting delays~--- helps overcome the practical limitation of producing out-of-range delay values with the readout function $R_{f_d}$. In this case, the values produced by $R_{f_d}$ are bounded between 0 and the maximum buffer size at the output ports of forwarding devices. Note that the buffer size is a device-specific feature that is independent of the network size. 

Lastly, RouteNet-F produces flow-level delay predictions $\hat{y}_{f_d}$ by combining the estimated queuing and transmission delays. The queuing delay~$\hat{d}_q$ is indirectly estimated by using the effective queue occupancies (in bits) on queues for a particular flow. Particularly, queue occupancy values are estimated by the readout function $R_{f_d}(\boldsymbol{h}^T_{f,l})$. Then, they are divided by the capacity of the link connected to the output port~$x_{l_c}$ to eventually produce a queuing delay estimate~$\hat{d}_q$. Likewise, the transmission delay~$\hat{d}_t$ is computed by dividing the mean flow packet size $x_{f_{ps}}$ by the link capacity $x_{l_c}$. With this, RouteNet-F estimates the delay of a flow after passing through a specific forwarding device and a link ($\hat{d}_{link}$):

\begin{align}
\hat{d}_q = \frac{R_{f_d}(\boldsymbol{h}^T_{f,l})}{x_{l_c}}\label{eq:queuing-delay}\\
\hat{d}_t = \frac{x_{f_{ps}}}{x_{l_c}} \label{eq:transmission-delay}\\
\hat{d}_{link} = \hat{d}_q + \hat{d}_t \label{eq:final-delay}
\end{align}

Hence, we can compute end-to-end flow delays as the sum of all the link delays $\hat{d}_{link}$ along the flow (loop lines~\mbox{\ref{readout}-\ref{lin:readout-end}} in Algorithm~\ref{alg:architecure}). Note that the function responsible for computing the effective queue occupancy, $R_{f_d}(\boldsymbol{h}^T_{f,l}$) takes as input the hidden state of the flow at a specific link $\boldsymbol{h}^T_{f,l}$, instead of directly considering queue states $\boldsymbol{h}^T_{q}$. This is because each flow can experience a different queuing behavior depending on its properties (e.g., traffic burstiness). 

Likewise, jitter estimates $\hat{y}_{f_j}$ are produced by combining the jitter predictions of all links along the flow. These predictions are made by the $ R_{f_j}$ function, which takes as input the hidden state of the flow at a specific link $\boldsymbol{h}^T_{f,l}$. Note that we define the jitter as the relative fluctuation with respect to the mean delay, that is, the ratio between the delay variance divided by the flow mean delay.

In the case of packet loss, RouteNet-F makes predictions directly on flows' hidden states $h_f^T$. We define the packet loss as the relative ratio of packets dropped with respect to the packets transmitted by the source; hence it is a bounded value $\hat{y}_{f_l}\in[0,1]$. We estimate it with the $ R_{f_l}$ function (line~\ref{r-loss} in Algorithm~\ref{alg:architecure}).

\subsection{Training and Implementation}

We implement RouteNet-Fermi in TensorFlow and it is publicly available at~\cite{github-fermi}.

As in any other ML model, fine-tuning the hyperparameters of RouteNet-F is crucial for achieving optimal performance and accuracy. Two key parameters to consider are the hidden state vectors ($\boldsymbol{h}_f$, $\boldsymbol{h}_q$, $\boldsymbol{h}_l$) and the number of message passing iterations ($T$). The size of the hidden state vectors determines how much information the model can encode, with larger sizes allowing for more information but also harming performance. Similarly, a larger number of message-passing iterations can help the model reach a higher level of convergence, but at the cost of increased computation. Through grid search experiments, we selected a set of hyperparameters that provide good accuracy while also being efficient. Specifically, we set the size of all the hidden state vectors to 32 elements and $T$ to 8 message-passing iterations.

We implement the functions $FRNN$ (Flow-Level RNN), $LRNN$ (Link-Level RNN), and $U_q$ (Queue Update Function) as Gated Recurrent Units (GRU). Functions $HS_f$, $HS_q$, and $HS_l$ are implemented as 2-layer fully-connected neural networks with ReLU activation functions with 32 units each. Similarly, functions $R_{f_d}$, $R_{f_j}$, and $R_{f_l}$ are implemented as a 3-layer fully-connected neural networks with ReLU activation function for the hidden layers, and a linear one for the output layer (except for $R_{f_l}$ that uses a Sigmoid activation function). Note that, the whole architecture of RouteNet-F (Algorithm~\ref{alg:architecure}) constitutes a fully-differentiable function. This means that it can be trained end-to-end using as input the network samples and as output the different flow-level performance metrics (e.g., delay, jitter, packet loss) as illustrated in the black box diagram of Figure~\ref{fig:RouteNet-F-black-box}.

For each set of experiments, we train RouteNet-F during 150 epochs of 2,000 samples each. We set as loss function the Mean Absolute Percentage Error for the delay experiments, the Mean Squared Error for jitter, and the Mean Absolute Error for the packet loss. In all the cases, we use an Adam optimizer with an initial learning rate of 0.001.

\section{Evaluation}\label{sec:evaluation}

In this section, we evaluate the performance of \mbox{RouteNet-F} in a wide range of relevant scenarios. First, we evaluate RouteNet-F in a variety of scenarios with complex traffic models and scheduling policies and compare it to a state-of-the-art Queuing Theory (QT) benchmark from~\cite{ferriol2022routenet}. In it, the network is modeled as a $M/M/1/b$ system where each queue along a path is treated independently. Note that the baseline, like the majority of QT models, assumes that arrivals to each queue are approximated by a Poisson process and that service times are exponentially distributed. Under these assumptions, the model derives analytical results for queue throughput, delay distributions, and blocking probabilities. A public implementation of the QT model can be found at~\cite{github-qt}.

Then, we evaluate RouteNet-F's generalization capabilities when evaluated in topologies x30 times larger than the ones seen during training and compare its inference times. Finally, we benchmark RouteNet-F in real-world scenarios with data from a real testbed, with traffic coming from real-world networks, and compare its performance with MimicNet~\cite{zhang2021mimicnet} a state-of-the-art DL-based model.

\subsection{Performance Analysis}\label{subsec:perf_analysis}

\subsubsection{Methodology}

In all the experiments (except for subsection~\ref{subsubsec:testbed} that comes from a real testbed), the ground truth is obtained using a packet-level simulator (Section~\ref{subsubsec:dataset}). Unless specified, in each evaluation we perform 50k experiments with a random configuration (src-dst routing, traffic intensity, per-interface scheduling policy, queue size, and traffic model), and compute the mean average delay, jitter, and packet loss. Then we compute the error of RouteNet-F's and QT's estimates with respect to the results of the packet simulator. For a fair comparison, the evaluation samples have not been used in the training phase of RouteNet-F.

\subsubsection{Dataset}\label{subsubsec:dataset}

We generate our dataset by simulating a wide range of network scenarios with the OMNet++ network simulator, v5.5.1~\cite{varga2001discrete}. An image of the simulator is publicly available and can be found at~\cite{github-bnnetsim}. Each dataset sample corresponds to a single simulation, and we record the mean delay, jitter, and packet loss for all the flows in the network, as well as queue-level statistics(e.g., mean occupancy, average packet loss, or average packet size). We select the different scenarios to simulate by randomly sampling from the possible values of all the input variables (Table~\ref{tab:sim_variables}). The traffic models \textit{autocorrelated exponentials} and \textit{modulated exponentials} reproduce realistic Internet traffic~\cite{KaragiannisNonStationaruy2004, ferriol2022routenet}.
We define the traffic intensity ($TI$) as a tunable parameter that defines the overall traffic load in the network scenario. $TI$ represents how congested is the network. In our dataset, it ranges from 400 to 2000 bits per time unit, with 400 being the lowest congested network (0\% avg. packet loss) and 2000 a highly congested network ($\approx$3\% avg. packet loss).

\begin{table}[!b]
\caption{Simulation variables.}
\label{tab:sim_variables}
\centering
\resizebox{\columnwidth}{!}{%
\begin{tabular}{c|l}
\toprule
\multirow{4}{*}{Topology}     & NSFNET~\cite{hei2004wavelength}, GEANT~\cite{barreto2012fast},       \\
                         &    GBN~\cite{pedro2011performance}, and scale-free synthetic   \\
                         &   topologies following  the Power-    \\
                         & Law Out-Degree algorithm~\cite{palmer2000generating}.  \\
\midrule
\multirow{4}{*}{Traffic Model}  & 6 options:  Poisson, On-Off, Constant\\
                                &   Bitrate, Autocorrelated Exponentials,  \\
                                & Modulated Exponentials (according \\
&  to~\cite{ferriol2022routenet}), and all models mixed.    \\

\midrule
 \multirow{2}{*}{Traffic Intensity} &   Random traffic intensities to generate \\
                                    & packet loss between 0\% and 3\%.  \\
\midrule
\multirow{5}{*}{Queuing Configuration}    &  1, 2, or 3 queues per port.    \\   
                                        &   Queue size: 8, 16, 32, or 64 kbits. \\   
                                        &   Policy: First In First Out, Strict   \\
                                        & Priority, Weighted Fair Queuing,                   \\   
                                        & and Deficit Round Robin.                    \\   
\bottomrule
\end{tabular}%
}
\end{table}

\subsubsection{Traffic Models} \label{subsubsec:traffic_models}

\begin{table}[!htbp]
\caption{Delay prediction using the QT baseline and RouteNet-F for different traffic models. The error is computed w.r.t. simulation results.}
\label{tab:traffic_model_qt_rf_delay}
\centering
\resizebox{\columnwidth}{!}{%
\begin{tabular}{ccccccccc}
\toprule
     & \multicolumn{4}{c}{QT}                 & \multicolumn{4}{c}{RouteNet-F}                \\
     \cmidrule(lr){2-5}
 \cmidrule(lr){6-9}
     & MAPE & MSE & MAE & R\textsuperscript{2} & MAPE & MSE & MAE & R\textsuperscript{2} \\
     \midrule
Poisson & 12.6\% & \textbf{0.001} & \textbf{0.017} & 0.998 & \textbf{2.1\%} & \textbf{0.001} & \textbf{0.017} & \textbf{0.999} \\
Deterministic & 22.4\% & 0.715 & 0.321 & 0.611 & \textbf{4.43\%} & \textbf{0.029} & \textbf{0.048} & \textbf{0.984} \\
On-Off & 23.1\% & 0.784 & 0.363 & 0.613 & \textbf{2.90\%} & \textbf{0.009} & \textbf{0.035} & \textbf{0.995} \\
A. Exponentials & 21.1\% & 0.686 & 0.316 & 0.618 & \textbf{2.62\%} & \textbf{0.010} & \textbf{0.030} & \textbf{0.994} \\
M. Exponentials & 68.1\% & 1.10 & 0.798 & 0.145 & \textbf{5.21\%} & \textbf{0.013} & \textbf{0.061} & \textbf{0.989} \\
Mixed & 35.1\% & 0.721 & 0.430 & 0.560 & \textbf{4.71\%} & \textbf{0.018} & \textbf{0.054} & \textbf{0.988} \\
\bottomrule
\end{tabular}%
}
\end{table}

\begin{table}[!htbp]
\caption{Jitter prediction using the QT baseline and RouteNet-F for different traffic models. The error is computed w.r.t. simulation results.}
\label{tab:traffic_model_qt_rf_jitter}
\centering
\resizebox{\columnwidth}{!}{%
\begin{tabular}{ccccccccc}
\toprule
     & \multicolumn{4}{c}{QT}                 & \multicolumn{4}{c}{RouteNet-F}                \\
     \cmidrule(lr){2-5}
 \cmidrule(lr){6-9}
     & MAPE & MSE & MAE & R\textsuperscript{2} & MAPE & MSE  & MAE  & R\textsuperscript{2} \\
     \midrule
Poisson & 71.9\% & 0.013 & 0.072 & 0.849 & \textbf{6.26\%} & \textbf{0.001} & \textbf{0.013} & \textbf{0.980} \\
Deterministic & 99.0\% & 0.057 & 0.067 & -1.86 & \textbf{7.17\%} & \textbf{0.001} & \textbf{0.008} & \textbf{0.924} \\
On-Off & 69.4\% & 0.057 & 0.098 & 0.425 & \textbf{8.50\%} & \textbf{0.004} & \textbf{0.018} & \textbf{0.959} \\
A. Exponentials & 74.3\% & 0.025 & 0.067 & 0.246 & \textbf{6.29\%} & \textbf{0.001} & \textbf{0.008} & \textbf{0.973} \\
M. Exponentials & 91.4\% & 1.34 & 0.834 & -0.622 & \textbf{10.3\%} & \textbf{0.036} & \textbf{0.091} & \textbf{0.956} \\
Mixed & 69.1\% & 0.299 & 0.296 & 0.025 & \textbf{9.82\%} & \textbf{0.007} & \textbf{0.034} & \textbf{0.974} \\
\bottomrule
\end{tabular}%
}
\end{table}

\begin{table*}[!th]
\caption{Packet Loss evaluation - Mean Absolute Error and Coefficient of Determination (R\textsuperscript{2}) of QT and RouteNet-F for the different traffic models.}
\label{tab:loss_traffic_model}
\centering
\resizebox{\textwidth}{!}{%
\begin{tabular}{ccccccccccccc}
\toprule
     & \multicolumn{2}{c}{Poisson} & \multicolumn{2}{c}{Const. Bitrate} & \multicolumn{2}{c}{On-Off} & \multicolumn{2}{c}{A. Exponentials} & \multicolumn{2}{c}{M. Exponentials} & \multicolumn{2}{c}{Multiplexed} \\
     \cmidrule(lr){2-3} \cmidrule(lr){4-5} \cmidrule(lr){6-7} \cmidrule(lr){8-9}\cmidrule(lr){10-11}\cmidrule(lr){12-13}
     & MAE & R\textsuperscript{2} & MAE & R\textsuperscript{2} & MAE & R\textsuperscript{2} & MAE & R\textsuperscript{2} & MAE & R\textsuperscript{2} & MAE & R\textsuperscript{2} \\
     \midrule
QT & 1.0\% & 0.97 & 11\% & 0.64 & 9.5\% & 0.63 & 10\% & 0.65 & 9.5\% & 0.08 & 4.6\% & 0.50  \\
RouteNet-F & \textbf{0.3\%} & \textbf{0.99} & \textbf{1.0\%} & \textbf{0.99} & \textbf{1.0\%} & \textbf{0.99} & \textbf{1.2\%} & \textbf{0.99} & \textbf{1.1\%} & \textbf{0.98} & \textbf{0.50\%} & \textbf{0.99}  \\
\bottomrule
\end{tabular}%
}
\end{table*}

This section analyzes the accuracy of RouteNet-F in a wide range of traffic models. The experiment is organized such that, for each traffic model, we add a degree of complexity by changing its first and second-order statistics (i.e., variance and autocorrelation). We start with simple traffic models such as Poisson or Constant Bitrate and end by testing more complex models that are better approximations of traffic seen in Internet links.

Tables~\ref{tab:traffic_model_qt_rf_delay} and~\ref{tab:traffic_model_qt_rf_jitter} show the errors of the delay and jitter for both, RouteNet-F and QT, with respect to the values obtained using the simulator. We can see that RouteNet-F achieves excellent accuracy results, producing very accurate estimates of delay and jitter in all traffic models: the worst cases are 5.21\% and 10.40\% for delay and jitter, respectively.

As expected, the estimates of the QT model are unacceptable in continuous-state traffic models, e.g. almost 70\% for modulated exponentials. On the other hand, it achieves moderate accuracy for discrete-state models (Poisson, Deterministic, and On-Off). It is also noticeable how the QT model shows poor accuracy across all the jitter estimations. The main reason for this is that QT assumes independence between queues in the network. Hence, the estimator used to compute the jitter is the sum of the individual delay variance of queues along the flow's paths.

For non-Markovian traffic models (e.g., On-Off), RouteNet-F produces accurate estimates, as well as for more challenging models that implement strong autocorrelation (Autocorrelated Exponentials) and heavy-tail distributions (Modulated Exponentials). These models are important since they approximate real traffic generated by  TCP~\cite{figueiredo2002autocorrelation}, similar to that found at Internet links~\cite{popoola2017empirical}. Also, notice that this traffic model could be made even more difficult for QT by increasing both the variance and the autocorrelation factor.

We add a final scenario (Mixed) where each src-dst pair generates traffic by randomly selecting one of the five available traffic models, and using random parameters for these models. In other words, we multiplex all traffic models in a single network topology. As the table shows, \mbox{RouteNet-F} shows good accuracy not only when modeling individual traffic models, but also when they are mixed across links in the network. It is worth noting that although RouteNet-F shows excellent accuracy for arbitrary parameterizations of these 6 traffic models, it cannot generalize to new traffic models not introduced during the training phase.

Finally, Table~\ref{tab:loss_traffic_model} shows the different metrics for the packet loss ratio. Since there are some paths where the packet loss ratio is zero, we provide the Mean Absolute Error and the Coefficient of Determination (R\textsuperscript{2}). The packet loss ratio is measured as the percentage of packets dropped w.r.t. packets sent, that is why the MAE is expressed in \% units. In this particular case, it is noticeable how the QT baseline works well in the scenario with the Poisson traffic model. However, the accuracy decreases remarkably in more complex scenarios. On the other hand,  RouteNet-F obtains a high accuracy with a worst-case MAE of 1.2\% and R\textsuperscript{2}$\geq$0.98.

\subsubsection{Scheduling} \label{subsec:scheduling}

This section aims to validate if RouteNet-F is capable of modeling the behavior of queues in the presence of several scheduling policies. For this purpose, we train the model using samples with mixed queue scheduling policies across nodes in the GEANT and NSFNET topologies. Then, we evaluate the model on samples of the GBN topology (unseen during training). In this experiment, each router port implements three different queues with a randomly selected scheduling policy for the queues: $(i)$ First In, First Out (FIFO), $(ii)$ Weighted Fair Queueing (WFQ), $(iii)$ Deficit Round Robin (DRR), and $(iv)$ Strict Priority (SP). For WFQ and DRR, the set of weights is also randomly selected. Furthermore, each flow has been assigned a Quality-of-Service class that maps it to a specific queue depending on the flow priority. To provide a fair benchmark with QT, in this experiment we use only Poisson traffic.

\begin{table}[!htbp]
\caption{Delay and jitter evaluation - Mean Absolute Percentage Error of QT and RouteNet-F in the presence of Scheduling Policies for low, medium, and high traffic intensity.}
\label{tab:scheduling_delay_jitter}
\centering
\resizebox{\columnwidth}{!}{%
\begin{tabular}{ccccccccc}
\toprule
     & \multicolumn{3}{c}{Delay}                 & \multicolumn{3}{c}{Jitter}                \\
     \cmidrule(lr){2-4}
 \cmidrule(lr){5-7}
     & Low & Medium  & High  & Low & Medium  & High \\
     \midrule
QT & 13.0\% & 17.3\% & 25.1\% & 49.0\% & 53.2\% & 59.6\%  \\
RouteNet-F & \textbf{0.80\%} & \textbf{2.60\%} & \textbf{7.31\%} & \textbf{3.95\%} & \textbf{5.77\%} & \textbf{14.8\%}  \\
\bottomrule
\end{tabular}%
}
\end{table}

Table~\ref{tab:scheduling_delay_jitter} shows the Mean Average Percentage Error (MAPE) of the delay and jitter for three different traffic intensities: from low-loaded to highly-congested scenarios. According to~\cite{floyd2001difficulties} the average packet loss on the Internet is around 2\%-3\%. Based on this, in the highly-congested scenarios, the mean packet loss rate is around 3\% which we believe represents a wide range of realistic network scenarios.  We can see that RouteNet-F outperforms the QT benchmark in both metrics (delay and jitter), obtaining a MAPE of 3.57\% for delay and 8.17\% for jitter after averaging the results over the three traffic intensities.

Similarly, Table~\ref{tab:scheduling_loss} presents the results for packet loss. Again, RouteNet-F outperforms the QT benchmark showing an MAE of 0.7\% and an average R\textsuperscript{2} close to 0.99.

\begin{table}[!htbp]
\caption{Packet loss evaluation - Mean Absolute Error and Coefficient of Determination (R\textsuperscript{2}) of QT and RouteNet-F in the presence of Scheduling Policies for low, medium, and high traffic intensity.}
\label{tab:scheduling_loss}
\centering
\resizebox{\columnwidth}{!}{%
\begin{tabular}{ccccccccc}
\toprule
     & \multicolumn{2}{c}{Low} & \multicolumn{2}{c}{Medium} & \multicolumn{2}{c}{High}                \\
     \cmidrule(lr){2-3} \cmidrule(lr){4-5} \cmidrule(lr){6-7}
     & MAE & R\textsuperscript{2}  & MAE & R\textsuperscript{2} & MAE & R\textsuperscript{2} \\
     \midrule
QT & 4.55\% & -0.05 & 9.22\% & 0.00 & 10.6\% & 0.29  \\
RouteNet-F & \textbf{0.2\%} & \textbf{0.97} & \textbf{0.10\%} & \textbf{0.99} & \textbf{0.40\%} & \textbf{0.99}  \\
\bottomrule
\end{tabular}%
}
\end{table}

\subsection{Generalization and Scalability}

\subsubsection{Generalization to larger networks}
\label{subsubsec:generalize_larger}

\begin{figure}[!t]
\centering
\includegraphics[width=0.9\columnwidth]{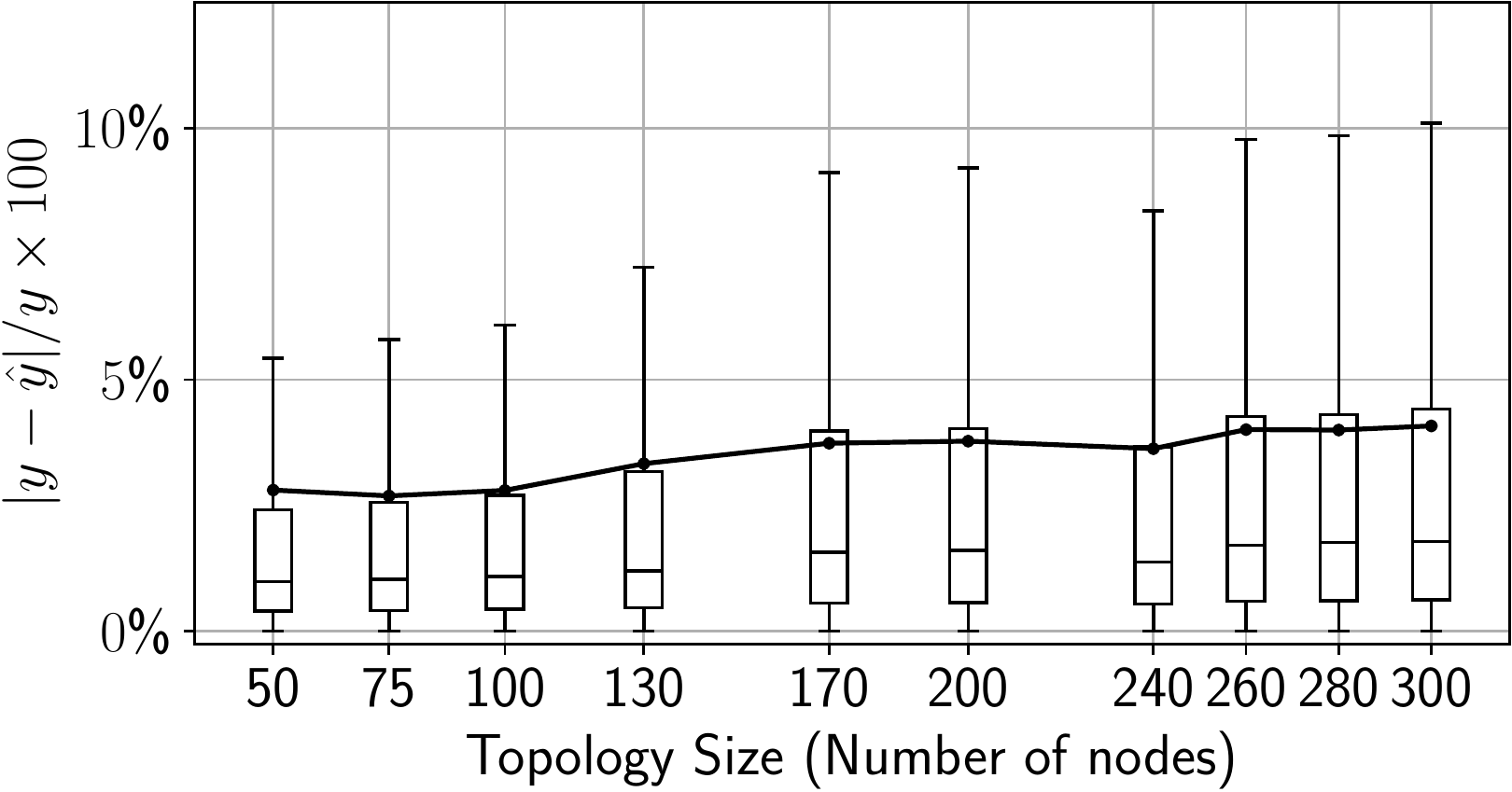}
\caption{Scaling with mixed traffic models and scheduling policies - Mean Absolute Relative Error of delay predictions vs. topology size, including different traffic models and queue scheduling configurations. The model was trained on a dataset with 10,000 samples from networks of 5 to 10 nodes.}
\label{fig:complete}
\end{figure}

The previous experiments show that RouteNet-F achieves remarkable accuracy when tested in scenarios with different traffic models (Section~\ref{subsubsec:traffic_models}) and different scheduling policies (Section~\ref{subsec:scheduling}) with topologies never seen in training. As previously discussed in Section~\ref{sec:challenges}, data-driven network models must generalize to larger networks than those seen during training to become a practical solution.

In this section, we evaluate RouteNet-F in a wide variety of networks significantly larger than the ones seen during the training phase. We generate a training set with 10,000 samples from networks of only 5 to 10 nodes. Following the process described in Sections~\ref{subsec:model} and~\ref{subsec:scheduling}, for each flow, we randomly assign a traffic model, and for each router port, we assign an arbitrary queue scheduling configuration. We evaluate the accuracy of RouteNet-F in topologies from 50 to 300 nodes, configuring the traffic models and the scheduling policies accordingly to the descriptions of Sections~\ref{subsec:model} and~\ref{subsec:scheduling}. In contrast to previous experiments, all these networks have been synthetically generated using the Power-Law Out-Degree algorithm described in~\cite{palmer2000generating}, where the ranges of $\alpha$ and $\beta$ parameters have been extrapolated from real-world topologies of the Internet Topology Zoo repository~\cite{knight2011internet}. Link capacities and generated traffic volumes are scaled accordingly.

Figure~\ref{fig:complete} shows the MAPE of the delay predictions made by RouteNet-F. We can observe that the proposed model obtains a worst-case error of $\approx$8\% for samples of networks with 300 nodes. This shows how RouteNet-F is capable of generalizing to networks 30x larger than those seen during training, even when introducing various traffic models and queue scheduling policies along the network. This is due to the capability of this model to effectively learn the underlying relationships between flows, links, and queues in the scenarios seen during training, and the posterior ability to exploit this learned knowledge in new scenarios not seen before.

Note that in this and the previous section, we have not tested any other baseline (e.g. RNN, QT) since they already fail in other \emph{relevant} scenarios.

Despite generalization is an open challenge in the field of Deep Learning (as previously discussed in Section~\ref{sec:limitations}) by using a custom GNN-based architecture and domain expert knowledge, RouteNet-F shows strong capabilities to generalize to considerably larger networks than the ones seen during training.

\subsubsection{Few-shot Learning}

\label{subsec:fewshot}
\begin{figure}[!t]
\centering
\includegraphics[width=0.9\columnwidth]{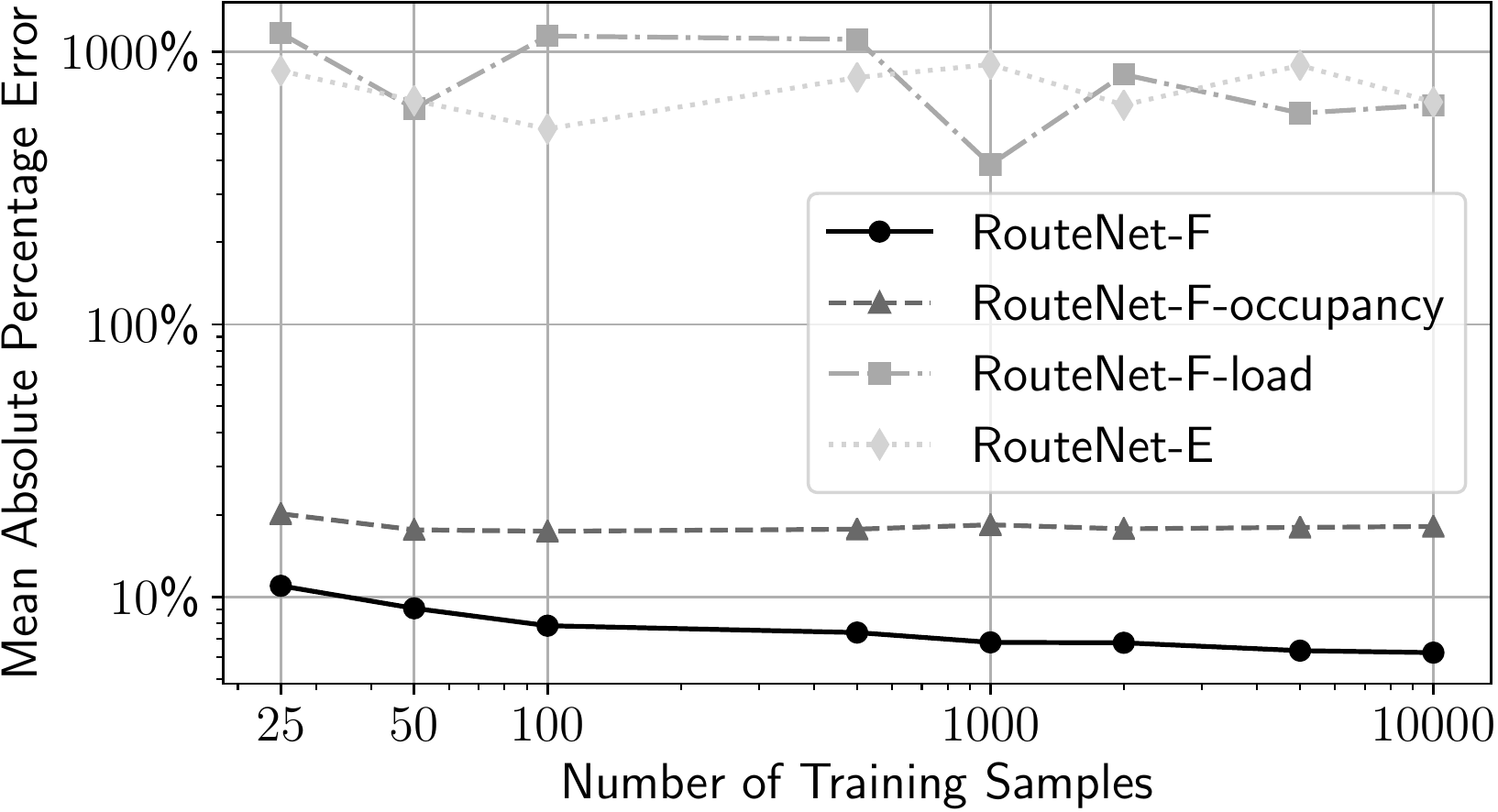}
\caption{Delay evaluation - Mean Absolute Percentage Error vs Number of Training Samples for the different versions of RouteNet-F and RouteNet-E.}
\label{fig:few_shot}
\end{figure}

The performance of DL models is often determined by the quantity and quality of the training data used. Here, it is important to consider that the process of collecting and labeling large amounts of data can be very costly and, sometimes, infeasible. In the field of computer networks, such data collection implies generating and storing data from costly network infrastructures. This can be generally a very expensive and time-consuming process. An alternative is the use of packet-level simulators, which are also very expensive in terms of computational cost. In this section, we evaluate the accuracy of RouteNet-F when trained with a very limited number of samples (i.e., few-shot learning). This may be very helpful to dramatically reduce the cost of generating the datasets and reduce the carbon footprint of training the model.

To do so, we train the model by randomly selecting 25, 50, 100, 1,000, 2,000, 5,000, and 10,000 samples from the previous training dataset (Section~\ref{subsubsec:generalize_larger}). Note that the topologies used for training range from 5 to 10 nodes, while the evaluation is done over samples of networks from 50 to 300 nodes. Figure~\ref{fig:few_shot} shows the MAPE with respect to the number of training samples used (see only results of \textit{RouteNet-F}). Interestingly, when trained with only 25 samples, the model shows an average error of 11\%. As the number of samples increases, RouteNet-F obtains slightly better accuracy, achieving an error of 6.24\% when trained with 10,000 samples.

\subsubsection{Ablation test}\label{subsubsec:ablation}

We aim to analyze which features of \mbox{RouteNet-F} have more impact on its accuracy. For this purpose, we perform an ablation test, by considering four models where we remove different features. The first one (labeled as \textit{RouteNet-F}) is the complete model used in the previous experiments, as it is previously described in Section~\ref{sec:routenet_fermi}. Second, in \mbox{\textit{RouteNet-F-occupancy}} we remove the link load as an input feature (see Sec.~\ref{sec:link-cap}), and replace it directly with the link capacity value as an initial feature of links $\boldsymbol{x}_l$. Third, \mbox{\textit{RouteNet-F-load}} predicts the flow-level delay using directly the hidden state of flows ($\boldsymbol{h}_{f}$), instead of predicting the effective queue occupancy of flows at a specific link ($\boldsymbol{h}_{f,l}$) and then adding up all the estimated link-level delays (see Sec.~\ref{subsec:out-dist}). Finally, we use \textit{RouteNet-E}~\cite{ferriol2022routenet} as a reference, which is the previous version of \mbox{RouteNet-F}, without any of the aforementioned features. Figure~\ref{fig:few_shot} shows the results obtained in this experiment. We can see that predicting the delay as the sum of link-level delays along flows (\textit{RouteNet-F-occupancy}) seems to have the largest impact on the accuracy of the model, achieving an error of 17.63\%. In addition, the results suggest that using the link load as input instead of the capacity  (\textit{RouteNet-F-load}) does not have a significant impact on the model's accuracy. However, when we combine this feature with the one of \textit{RouteNet-F-occupancy} we see a slight improvement. Particularly, in \textit{RouteNet-F}, which implements the two features, we can see that the prediction error decreases to 6.24\%. 

In this experiment, it is observed that the previous version of the model (\textit{RouteNet-E}) exhibits poor accuracy. This is likely because RouteNet-E was not designed to support scalability levels of 30x larger networks. Additionally, RouteNet-E assumes uniform delays for all flows traversing the same queue and link. In contrast, RouteNet-F takes into consideration that flows may experience different delays when traversing the same queue and link, depending on their traffic model (e.g., traffic burstiness).

\subsubsection{Scalability: Training and Inference time}\label{subsubsec:time}

Models that are capable of producing fast estimations are particularly interesting for network control and management applications, as they can be deployed in real-time scenarios. In this section, we evaluate the inference time of both RouteNet-F and the QT baseline. To this end, we measure the inference times of the experiments of the general evaluation (Sec.~\ref{subsubsec:generalize_larger}). Table~\ref{tab:time} shows that both models operate in the order of milliseconds for topologies lower than 110 nodes. Particularly, QT performs better for smaller topologies ($<$50 nodes). However, as the size of the topology increases,  RouteNet-F is faster. 

Finally, note that one difference between analytical models (e.g., QT) and DL-based models (e.g., RouteNet-F) is that the last need a training process that may be taken into consideration when comparing these times. These training times depend greatly on various factors like the size of the training dataset, the hyperparameters, the used hardware, etc. In our particular case, the model that obtained the higher accuracy was trained during 20 epochs of 2,500 samples each, lasting for about 2h30min.

In this experiment, we used one CPU [AMD Ryzen 9 3950X @ 3.5 GHz]. However, an advantage of DL-based models is that they can be easily parallelizable using hardware-specific solutions (e.g., GPU), thus reducing considerably their execution times in production.

\begin{table}[!t]
\caption{Inference time vs. topology size for RouteNet-F and the QT baseline.}
\label{tab:time}
\centering
\resizebox{\columnwidth}{!}{%
\begin{tabular}{ccccccccc}
\toprule
     & \multicolumn{6}{c}{Topology Size} \\
     \cmidrule(lr){2-2} \cmidrule(lr){3-3} \cmidrule(lr){4-4} \cmidrule(lr){5-5} \cmidrule(lr){6-6} \cmidrule(lr){7-7} 
     & 10 & 30 & 50 & 70 & 90 & 110 \\
     \midrule
RouteNet-F & 48.03 ms & 76.25 ms & 110.5 ms & 285.6 ms & 455.3 ms & 613.3 ms  \\
QT & 28.01 ms & 49.19 ms & 131.68 ms & 326.5 ms & 662.15 ms & 962.5ms   \\
\bottomrule
\end{tabular}%
}
\end{table}

\subsection{Benchmarking of RouteNet-F}

\subsubsection{Testbed} \label{subsubsec:testbed}

\begin{figure}[!ht]
\centerline{\includegraphics[width=0.9\columnwidth]{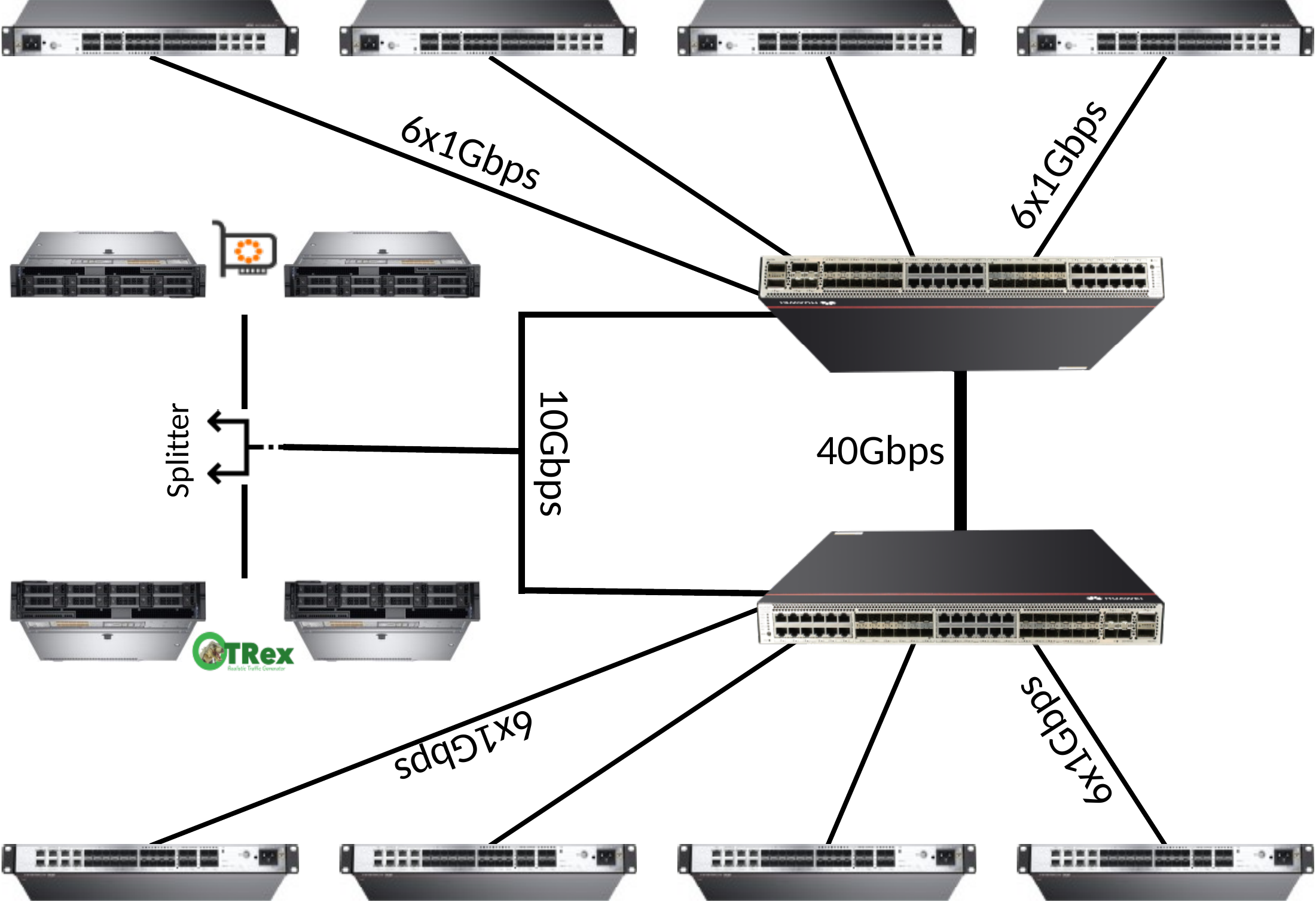}}
\caption{Schematic representation of the network testbed.}
\label{fig:testbed}
\end{figure}

In previous sections, we examined how RouteNet-F can be applied in different network scenarios. This section evaluates its performance in a real-world scenario using real-world hardware and synthetic traffic. For this, we set up a physical network testbed, as shown in Figure~\ref{fig:testbed}. This testbed includes $(i)$~8 Huawei NetEngine 8000 M1A routers, $(ii)$~2 Huawei S5732-H48UM 2CC 5G Bundle switches, and $(iii)$~4 servers. Two of the servers are used to generate traffic using the TRex traffic generator, and the other two are used for capturing and analyzing traffic with the PF\_RING software.

To train and test the model, we generated 1,000 samples with realistic network topologies (with a maximum of 8 nodes) and different routing, queueing, and traffic configurations. Out of these 1,000 samples, we randomly selected 800 samples for training the model and used the remaining 200 samples for testing. Note that, the algorithm~\ref{alg:architecure} described in Section~\ref{subsec:model} has been slightly modified to add the state of the queues of the switches found in the topology. This only affects the initialization process, and not the message-passing architecture itself.

Table~\ref{tab:testbed_and_real_traffic} shows the results of RouteNet-F. As can be seen, RouteNet-F obtains a remarkable performance (11\% MAPE) which is in line with the previous results when compared with simulated data.

\subsubsection{Real Traffic} \label{subsubsec:real_traffic}

In our previous experiments, we tested RouteNet-F using synthetic-simulated traffic. Now, we want to see how well this model performs when applied to actual traffic data. 

For this, we used real-world traffic data from the SNDlib library~\cite{SNDlib10} and combined it with packet inter-arrival times from a recent snapshot of the MAWI repository (Sample point 2022/09)~\cite{mawi}. We then scale the inter-arrival times to match the values in the traffic data. Additionally, we used a distribution of source-destination flows from a real internet service provider~\cite{schnitter2007quality} to map flows to different ToS classes. Our dataset includes 4 real-world network topologies, including one previously used (GEANT) and three new topologies that the model has never seen before (ABILENE, NOBEL-GBN, and GERMANY50). For this experiment, we leveraged the knowledge obtained from previous versions and used a previous checkpoint, fine-tuning it using 200 samples of the GEANT topology.

Table~\ref{tab:testbed_and_real_traffic} shows the results of RouteNet-F. We can see that RouteNet-F achieves a remarkable performance (5.67\% MAPE) when tested using real-world traffic data. Again, these results are close to the simulator and testbed ones.

\begin{table}[!htbp]
\caption{Delay prediction using RouteNet-F for the testbed and the real traffic traces experiments.}
\label{tab:testbed_and_real_traffic}
\centering
\resizebox{0.75\columnwidth}{!}{%
\begin{tabular}{ccccccccc}
\toprule
 \cmidrule(lr){6-9}
     & MAPE & MSE  & MAE  & R\textsuperscript{2} \\
     \midrule
Testbed & 11.0\% & 6.12x10\textsuperscript{-5} & 0.0007 & 0.869 \\
Real Traffic & 5.67\% & 1.66x10\textsuperscript{-5} & 0.0003 & 0.877 \\
\bottomrule
\end{tabular}%
}
\end{table}

\subsubsection{State-of-the-Art}

This section aims to compare the accuracy of RouteNet-F against MimicNet~\cite{zhang2021mimicnet}. MimicNet is a DL-based model which combines discrete event simulators with Deep Neural Networks (DNN). MimicNet takes advantage of the accuracy of discrete packet-event simulators to generate data of a small network which then is used to train an RNN-based estimator known as a "mimic". Finally, MimicNet composes several of those mimics to perform predictions of much bigger larger networks. This makes MimicNet an alternative to a discrete packet-event simulator as it reduces the cost of simulation for large networks, providing an accurate estimation of the per-packet level distributions. However, as stated in the MimicNet paper, the main limitation of this strategy is that it only works for FatTree topologies.

Following the parameters and the specifications described in~\cite{yangDeepQueueNet2022} we generate a dataset containing three different topologies (FatTree16, FatTree64, and FatTree128) and compute the average RTT.

Table~\ref{tab:fattree_mimic_rf_rtt} shows the results for both RouteNet-F and MimicNet. For RouteNet-F we show the same metrics as before and the Normalized Wasserstein Distance ($W_1$). Contrary to RouteNet, MimicNet does not directly compare the average RTT of the packets aggregated per path. Instead, it computes the distance of both distributions (the predicted and the real one).

Note that in this scenario, RouteNet-F achieves an outstanding accuracy not only when compared with MimicNet, but also when compared with a state-of-the-art QT model (Table~\ref{tab:traffic_model_qt_rf_delay}). This is mainly because, in previous experiments, we explored a wide variety of scenarios containing from low to high (3\%) packet loss ratios, while in this particular scenario, the packet loss rate is about 0.3\%.

Finally, as shown in section~\ref{subsubsec:time}, RouteNet-F is capable of predicting the performance metrics in the order of milliseconds. In contrast, MimicNet ranges from minutes for the smaller topologies to hours for the larger ones. This difference is mainly because, while RouteNet-F focuses on predicting the different performance metrics aggregated by flows, MimicNet predicts those metrics at a packet level.

\begin{table}[!ht]
\caption{Average RTT prediction using MimicNet and RouteNet-F for different FatTree topologies. The error is computed w.r.t. simulation results.}
\label{tab:fattree_mimic_rf_rtt}
\centering
\resizebox{\columnwidth}{!}{%
\begin{tabular}{ccccccccc}
\toprule
     & \multicolumn{1}{c}{MimicNet}                 & \multicolumn{5}{c}{RouteNet-F}                \\
     \cmidrule(lr){2-2}
 \cmidrule(lr){3-7}
     & $W_1$ & MAPE & MSE  & MAE  & R\textsuperscript{2} & $W_1$\\
     \midrule
FatTree16 & 0.0080 & \textbf{0.37\%} & \textbf{1.31x10\textsuperscript{-10}} & \textbf{5.33x10\textsuperscript{-6}} & \textbf{0.999} & \textbf{0.0018} \\
FatTree64 & 0.0122 & \textbf{0.44\%} & \textbf{1.70x10\textsuperscript{-10}} & \textbf{7.15x10\textsuperscript{-6}} & \textbf{0.999} & \textbf{0.0026} \\
FatTree128 & 0.0172 & \textbf{0.67\%} & \textbf{3.40x10\textsuperscript{-10}} & \textbf{1.20x10\textsuperscript{-5}} & \textbf{0.998} & \textbf{0.0060} \\
\bottomrule
\end{tabular}%
}
\end{table}

\section{Related Work}

The use of Deep Learning (DL) for network modeling has recently attracted much interest from the networking community. The authors from~\cite{wang2017machine} survey different techniques and discuss data-driven models that can learn from real networks. Initial attempts to implement this idea use fully-connected neural networks (e.g.,~\cite{valadarsky2017learning, mestres2018understanding}). Such early attempts do not generalize to different networks unseen in training, they are not tested with realistic traffic models, and they do not model QoS-aware queue scheduling configurations. More recent works propose elaborated neural networks models, like Variational Auto-encoders~\cite{xiao2018deep} or ConvNN~\cite{chen2018deep}. However, they have similar limitations. 

Since the introduction of Graph Neural Networks~\cite{scarselli2008graph}, they have already been applied to different fields such as chemistry~\cite{gilmer2017neural} or logistics~\cite{kosasih2021machine}. In the field of computer networks, early pioneering works leverage GNNs~\cite{geyer2019deeptma, rusek2019unveiling}. However, they use a basic GNN architecture that considers a simplified model of the network, ignoring traffic models, queuing policies, and the critical property of generalizing to larger networks. 

More recent works leverage GNNs to model complex characteristics of computer networks. For example, xNet~\cite{wang2022xnet} learns the state transition function between time steps and rolls it out to obtain the complete fine-grained prediction trajectory, but it does not take into account scaling to larger networks. Others focus on data center networks to predict the Flow Completion Time~\cite{li2020traffic}. Finally, state-of-the-art simulators MimicNet~\cite{zhang2021mimicnet} and DeepQueueNet~\cite{yangDeepQueueNet2022} leverage DL models (LSTMs and transformers, respectively) to accelerate parts of the simulation process, with a focus on per-packet delay prediction, as opposed to our per-path average delay prediction.

\section{Conclusion}

In this paper, we have presented RouteNet-Fermi, a custom GNN model designed for network performance analysis. This model supports a wide range of configuration parameters related to routing, queue scheduling, and traffic models while being able to accurately model networks 30 times larger than the ones seen during training. In our evaluation, we have shown that the proposed model outperforms a state-of-the-art queuing theory model, especially in scenarios with complex and realistic traffic models. At the same time, RouteNet-Fermi achieves comparable accuracy with respect to computationally-expensive packet-level simulators (MAPE $\approx$ 6.24\%) while exhibiting considerably lower inference times (on the order of milliseconds in networks of 100 nodes). Finally, we validated RouteNet-F in a wide variety of real-world scenarios including a testbed and real-world traffic traces.

\bibliographystyle{IEEEtran}
\bibliography{IEEEabrv,ton-references}

\end{document}